\documentclass[aps,pre,preprint,groupedaddress]{revtex4-1}

\usepackage{graphicx}
\usepackage{dcolumn}
\usepackage{amsmath}    
\usepackage{amssymb}
\usepackage{bm}
\usepackage{hyperref}
\usepackage{latexsym}
\usepackage{verbatim}
\usepackage{color}
\usepackage{subfigure}
\usepackage[autostyle]{csquotes}
\def\beq{\begin{equation}}
\def\eeq{\end{equation}}

\def\beq{\begin{equation}}
\def\eeq{\end{equation}}
\def\bea{\begin{eqnarray}}
\def\eea{\end{eqnarray}}

\begin{document}


\title{Subdiffusion, Anomalous Diffusion and Propagation of a Particle Moving in Random and Periodic Media}

\author{Shradha Mishra\footnote{present affiliation: Department of Physics, Indian Institute of Technology (BHU), Varanasi, UP 221005, India}}
\email[]{shradha.mishra@bose.res.in, smishra.phy@itbhu.ac.in}
\affiliation{Satyendra Nath Bose National Centre for Basic Sciences, Block-JD, Sector-III, Salt Lake, Kolkata-700098, India}
\author{Sanchari Bhattacharya}
\affiliation{Satyendra Nath Bose National Centre for Basic Sciences, Block-JD, Sector-III, Salt Lake, Kolkata-700098, India}
\author{Benjamin Webb}
\affiliation{Brigham Young University, Department of Mathematics, 308 TMCB, Provo, UT
84602, USA}
\author{E. G. D. Cohen}
\affiliation{The Rockefeller University, 1230 York Avenue, New York, New York 10021, USA}

\date{\today}

\begin{abstract}
We investigate the motion of a single particle moving on a two-dimensional square lattice whose sites are occupied by right and left rotators. These left and right rotators deterministically rotate the particle's velocity to the right or left, respectively and \emph{flip} orientation from right to left or from left to right after scattering the particle. We study three types of configurations of left and right rotators, which we think of as types of media, through with the particle moves. These are completely random (CR), random periodic (RP), and completely periodic (CP) configurations. For CR configurations the particle's dynamics depends on the ratio $r$ of right to left scatterers in the following way. For small $r\simeq0$, when the configuration is nearly homogeneous, the particle subdiffuses with an exponent of 2/3, similar to the diffusion of a macromolecule in a crowded environment. Also, the particle's trajectory has a fractal dimension of  $d_f\simeq4/3$, comparable to that of a self-avoiding walk. As the ratio increases to $r\simeq 1$, the particle's dynamics transitions from subdiffusion to anomalous diffusion with a fractal dimension of $d_f\simeq 7/4$, similar to that of a percolating cluster in 2-d. In RP configurations, which are more structured than CR configurations but also randomly generated, we find that the particle has the same statistic as in the CR case. In contrast, CP configurations, which are highly structured, typically will cause the particle to go through a transient stage of subdiffusion, which then abruptly changes to propagation. Interestingly, the subdiffusive stage has an exponent of approximately 2/3 and a fractal dimension of $d_f\simeq4/3$, similar to the case of CR and RP configurations for small $r$.
\end{abstract}
\maketitle



\section{Introduction}{\label{intro1}}
In a Lorentz \cite{Lorentz} lattice gas (LLG) a single particle moves along the bonds of a lattice. When  it arrives at a lattice site the particle encounters a scatterer, which scatters the particle according to some fixed rule. In addition to the particle, which has a position that changes over time, each scatterer can also have a number of different orientations, or more generally states, that may also change over time as it interact with the particle, etc.

In the past, there have been a number of ways in which a Lorentz \cite{Lorentz} lattice gas (LLG) has been used to model different physical phenomena \cite{langtonant}. The intent in most of these studies is to use LLG models to understand basic principles that underly dynamic processes such as diffusion, propagation, etc. Continuing in this manner, the goal of this paper is to better understand how the motion of a particle is effected when moving through a media that has little or no structure verses a media that has a high degree of regularity.

The \emph{media} or \emph{environment} through which the particle moves is modeled by the collective configuration of the scatterers. The initial orientation of each scatterer is called the LLG's \emph{initial configuration}. One of the main questions we investigate in this paper is how a particle, moving along the bonds of a lattice, will be effected by an initial configuration of scatterers, i.e. media, that is ordered or disordered to some degree.

For defining an LLG we need to choose (i) a lattice, (ii) a scattering rule, and (iii) an initial configuration or class of initial configurations. In previous studies, a wide variety of dynamics has been observed in such systems \cite{bunimovich, cao, grosfils, ruijgork, wang, wangcohen, wangcohenphysica, wangcohenstatphys, ben2014, ben2015} depending on the particular choice of (i)-(iii). In this article, we choose (i) the square lattice with (ii) flipping right and left rotators, which change orientation after scattering the particle. For (iii), the type of initial configurations we choose to study can be divided into three distinct categories. These we call completely random (CR), random periodic (RP), and completely periodic (CP) configurations, respectively. A detailed description of these configurations is given in section \ref{model}.



\begin{figure*}[htbp]
 \centering
 \subfigure[]{
   \includegraphics[scale=0.35]{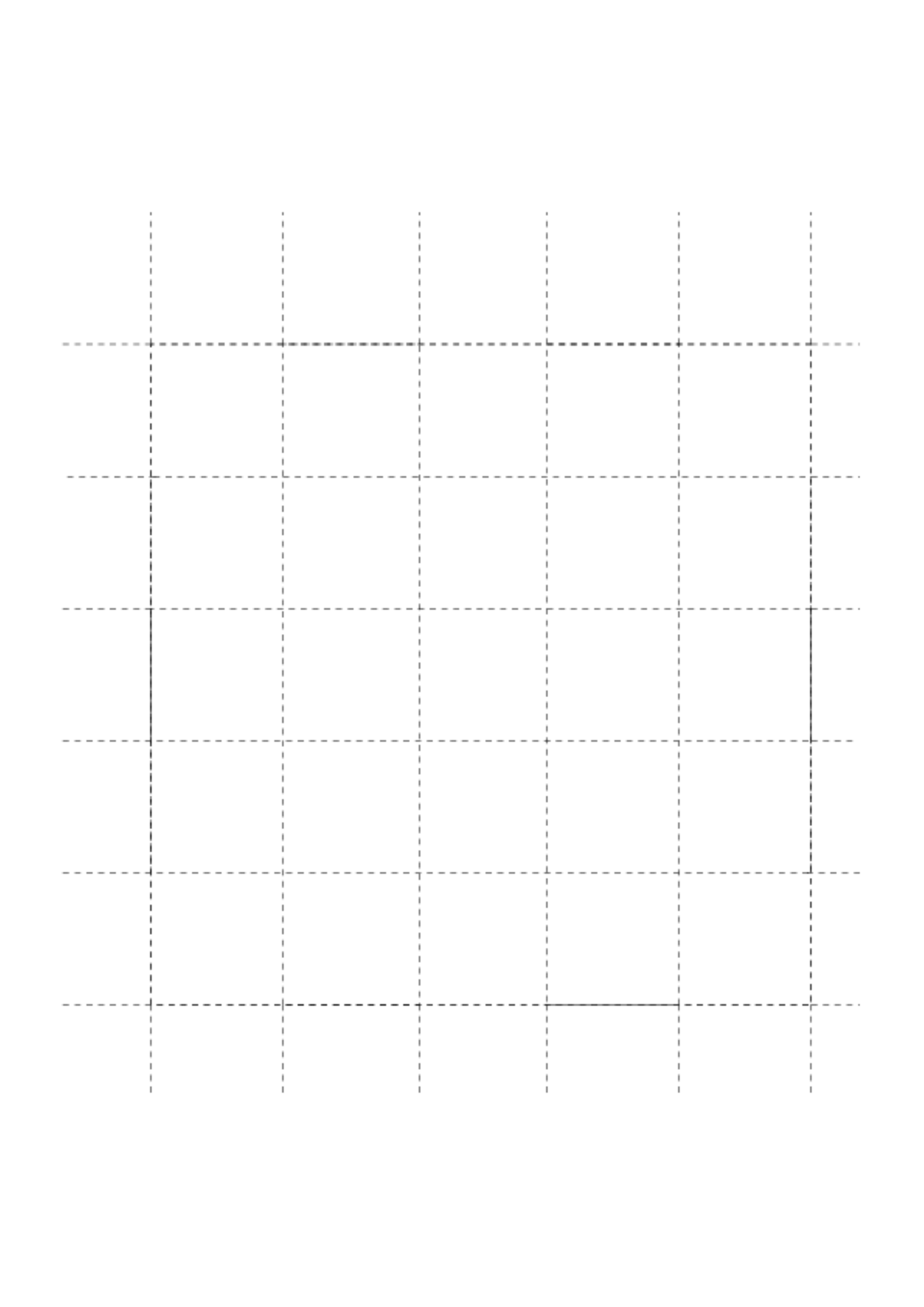}
    \label{fig:subfig1}
    }
   \subfigure[]{
     \includegraphics[scale=0.35]{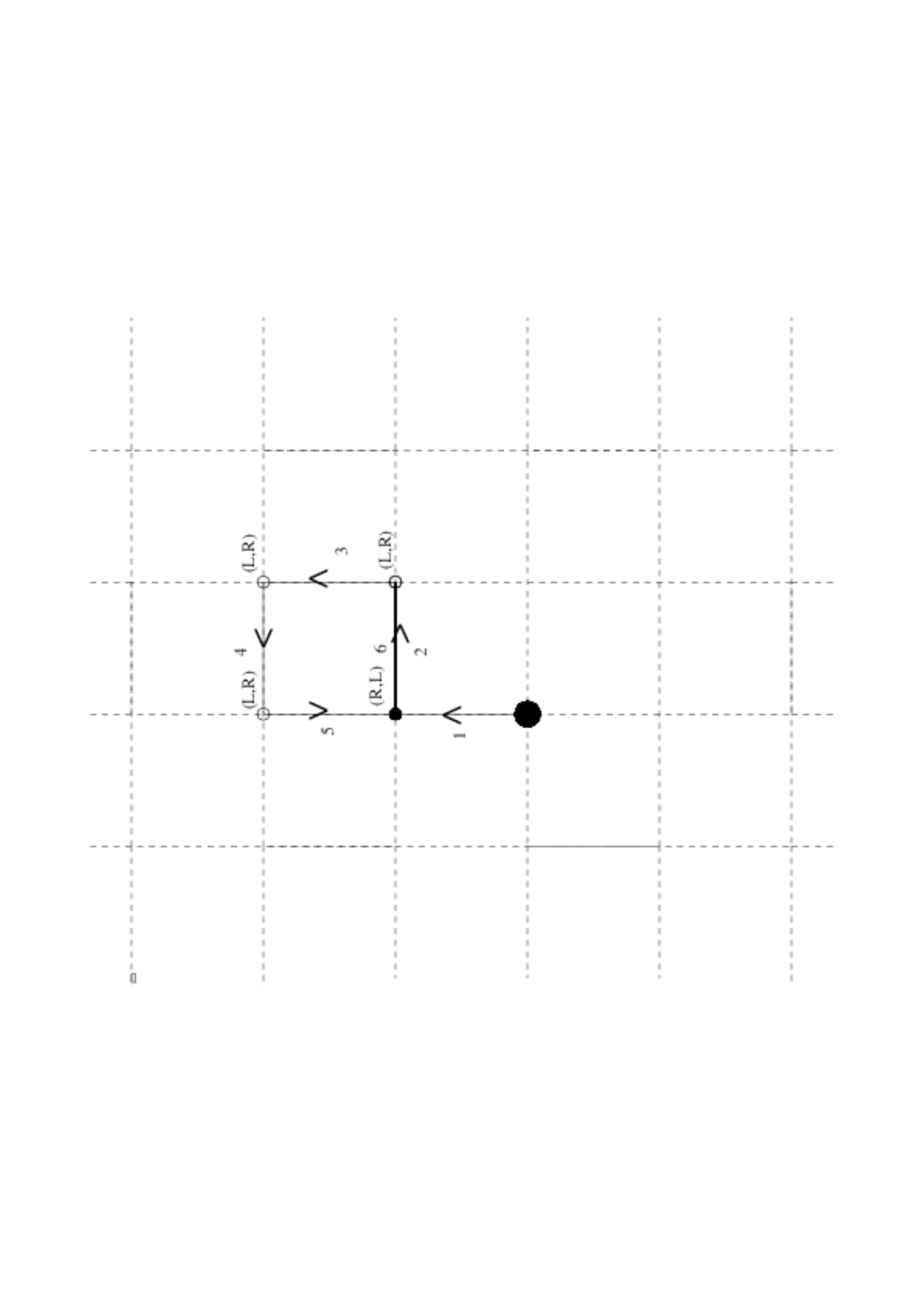}
      \label{fig:subfig2}
      }
\subfigure[]{
  \includegraphics[scale=0.28]{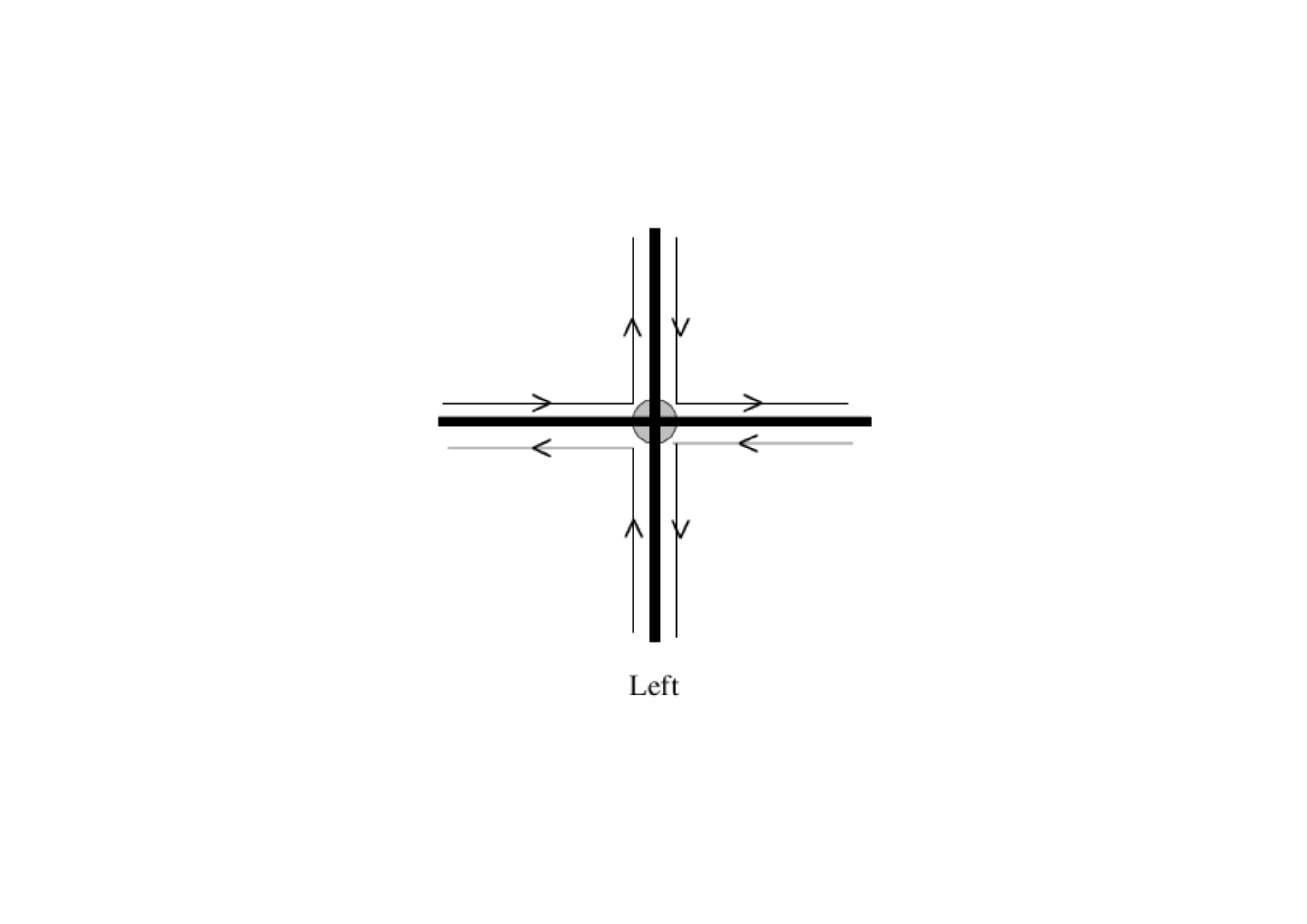}
   \label{fig:subfig3}
   }
 \subfigure[]{
  \includegraphics[scale=0.28]{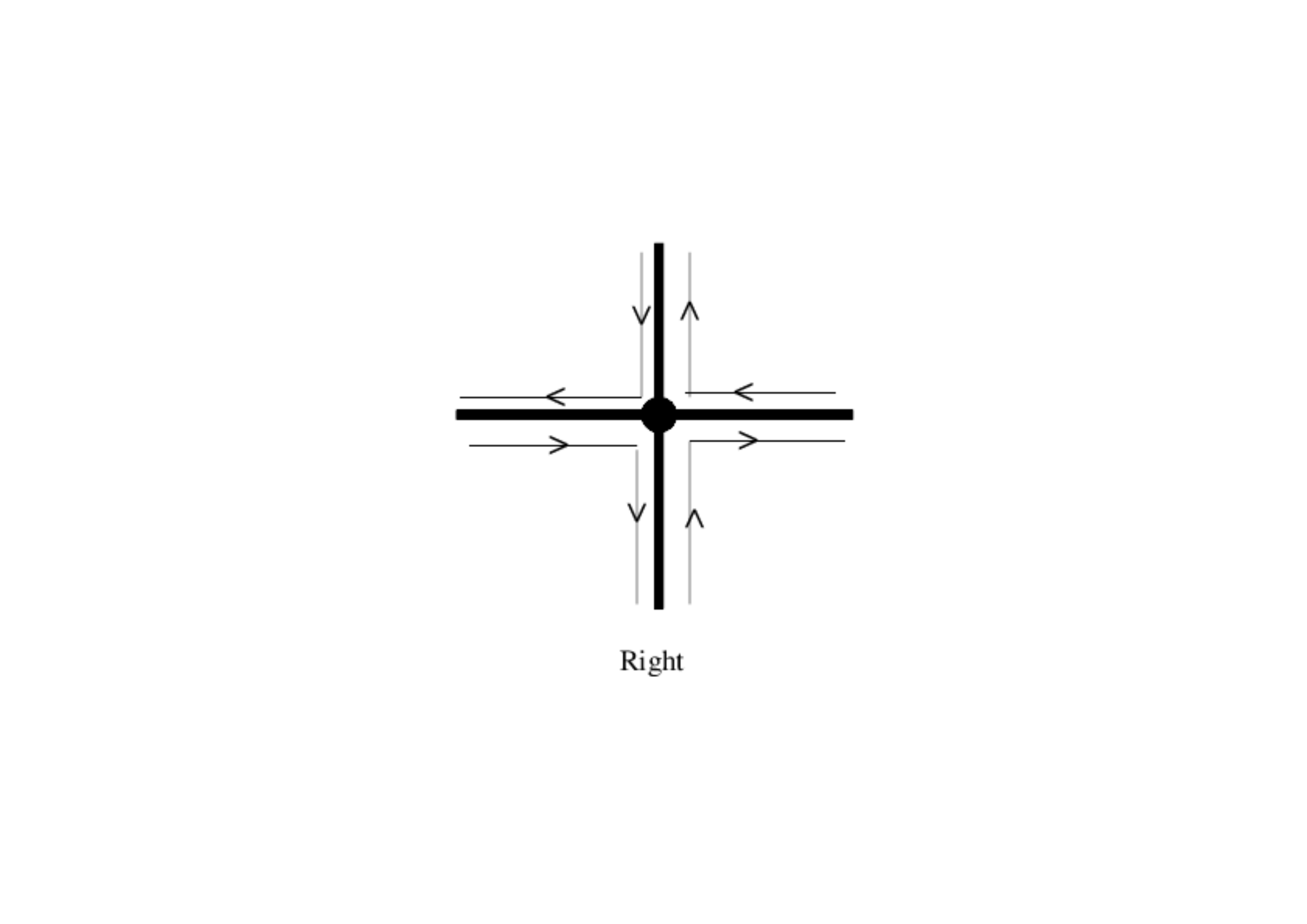}
   \label{fig:subfig4}
   }
\caption{A regular square lattice is shown in (a). In (b) a typical path of the particle is shown as it moves through the lattice. The large circle indicates the particle's initial position $\mathbf{r}(0)$, the small filled circles represent rotators that are initially oriented to the right, and the open circles indicate rotators that are initially oriented to the left. Arrows indicate the direction of particle's motion. The numbering 1-6 represents the sequence of lattice bonds the particle traverses as it moves through the lattice. The pair (X,Y) at the vertices indicate the rotator's initial orientation X at $t=0$ and its final orientation Y at $t=5$. The effect of a left and right rotator on the particle's motion is shown in (c) and (d), respectively. After scattering the particle to the right (left) the particle flips orientation becoming a left (right) rotator.}
\label{fig1}
\end{figure*}

More generally, the configurations we consider can be divided into those with some form of periodicity and those without. Those configurations with some amount of periodicity are the RP and CP configurations. The particle's motion over these configurations can potentially be applied to model certain types of physical or biological processes \cite{Felix, horton}. The application we consider is to model the particle's motion in a media in which there are periodic defects. These defects can either be randomly distributed throughout the media with a certain probability in a periodic fashion (RP configurations) or distributed throughout the media with a fixed periodicity (CP configurations). Such defects can be thought of as \enquote{irregularities} in a particular material or media. Alternatively, in the biological sense we may think of these defects as \enquote{mutations} if there is some way in which these lattice defects \enquote{mutate} the particle's dynamics.

Besides periodic configurations, we also consider random configurations in which a right rotator is placed at a lattice site with probability $(r+1)^{-1}$ and a left rotator with probability $1-(r+1)^{-1}$ for some $r\in[0,\infty]$. The number $r=\frac{C_L}{C_R}$ is then the \emph{ratio} of left to right rotators on the lattice, where $C_R$ and $C_L$ are concentration of left and right rotators, respectively. For $r=0$ the configuration of rotators is completely \emph{homogenous} consisting of only right rotators, whereas $r=1$ corresponds to the case in which left and right rotators are evenly distributed throughout the lattice. For $r\simeq 0$ there is a small number of left rotators on the lattice, which we consider to be the collection of \emph{lattice defects}.

In this paper we study both the transient and asymptotic behavior of a particle moving on a two-dimensional square lattice occupied by different initial configurations of right and left rotators. The main goal is to describe in what way the random and periodic configurations CR, RP, and CP effect the particle's dynamics for various ratios $r$ of left to right rotators.

The paper is organized as follows. In section \ref{model} we formally describe both the LLG and the various types of random and/or periodic configurations we study in this paper. Section \ref{results} then gives numerical results regarding the various properties the particle exhibits for different initial configurations and ratios of left and right rotators. We conclude in section \ref{discussion} with a number of observations regarding these results and mention some related problems for future work.


\begin{figure}[htbp]
 \centering
 \subfigure[]{
  \includegraphics[scale=0.2]{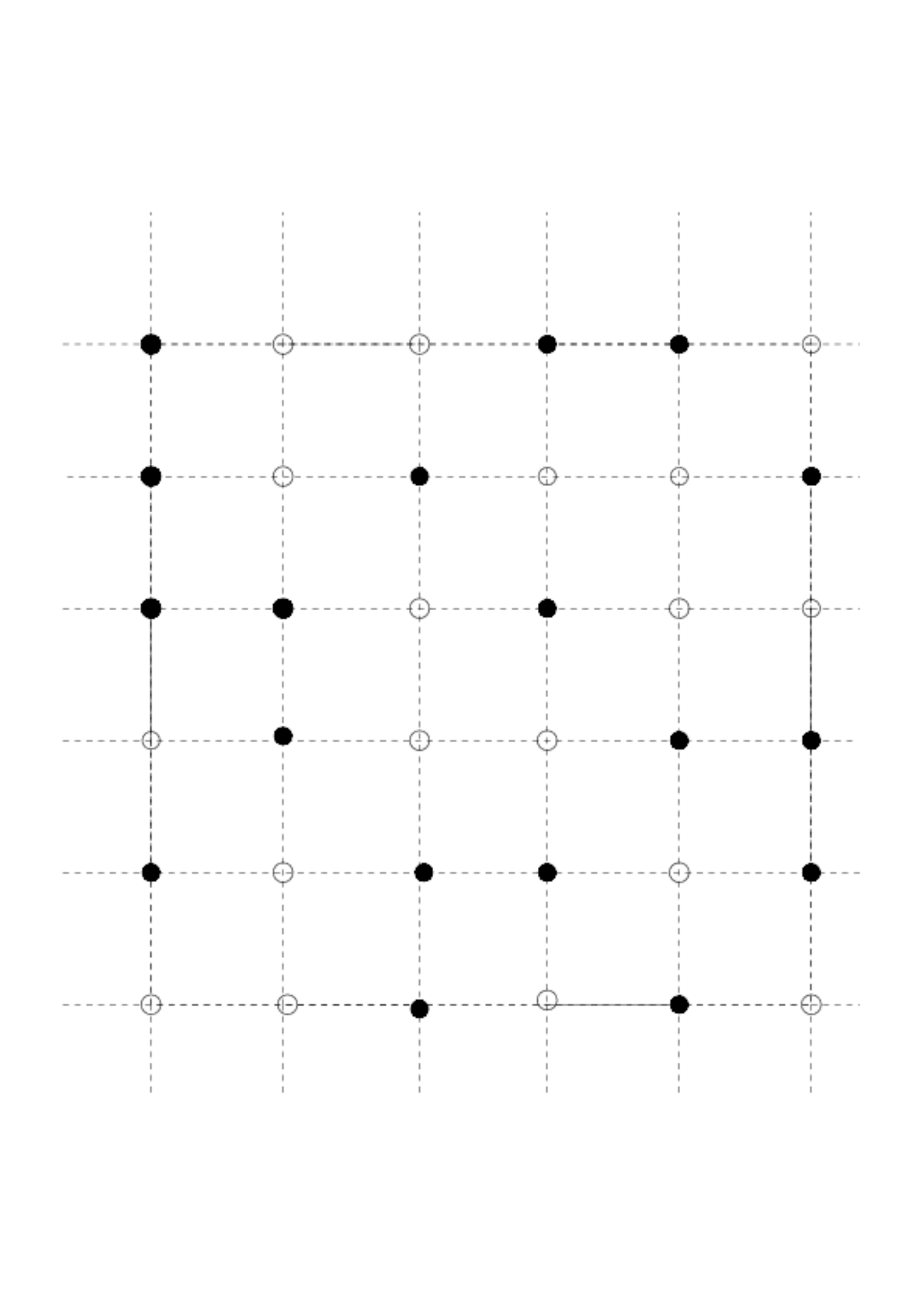}
   }
 \subfigure[]{
  \includegraphics[scale=0.2]{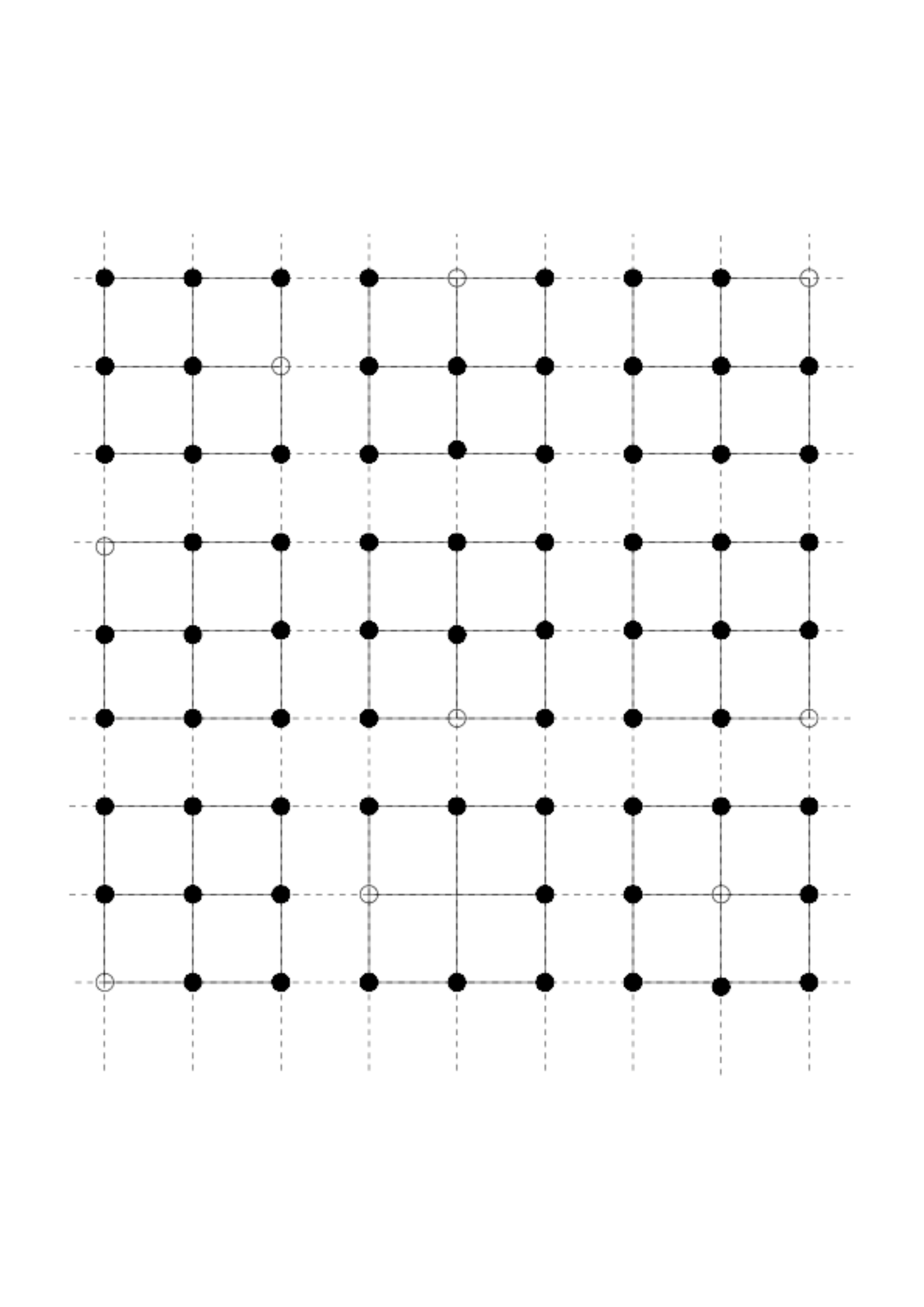}
   }
 \subfigure[]{
  \includegraphics[scale=0.2]{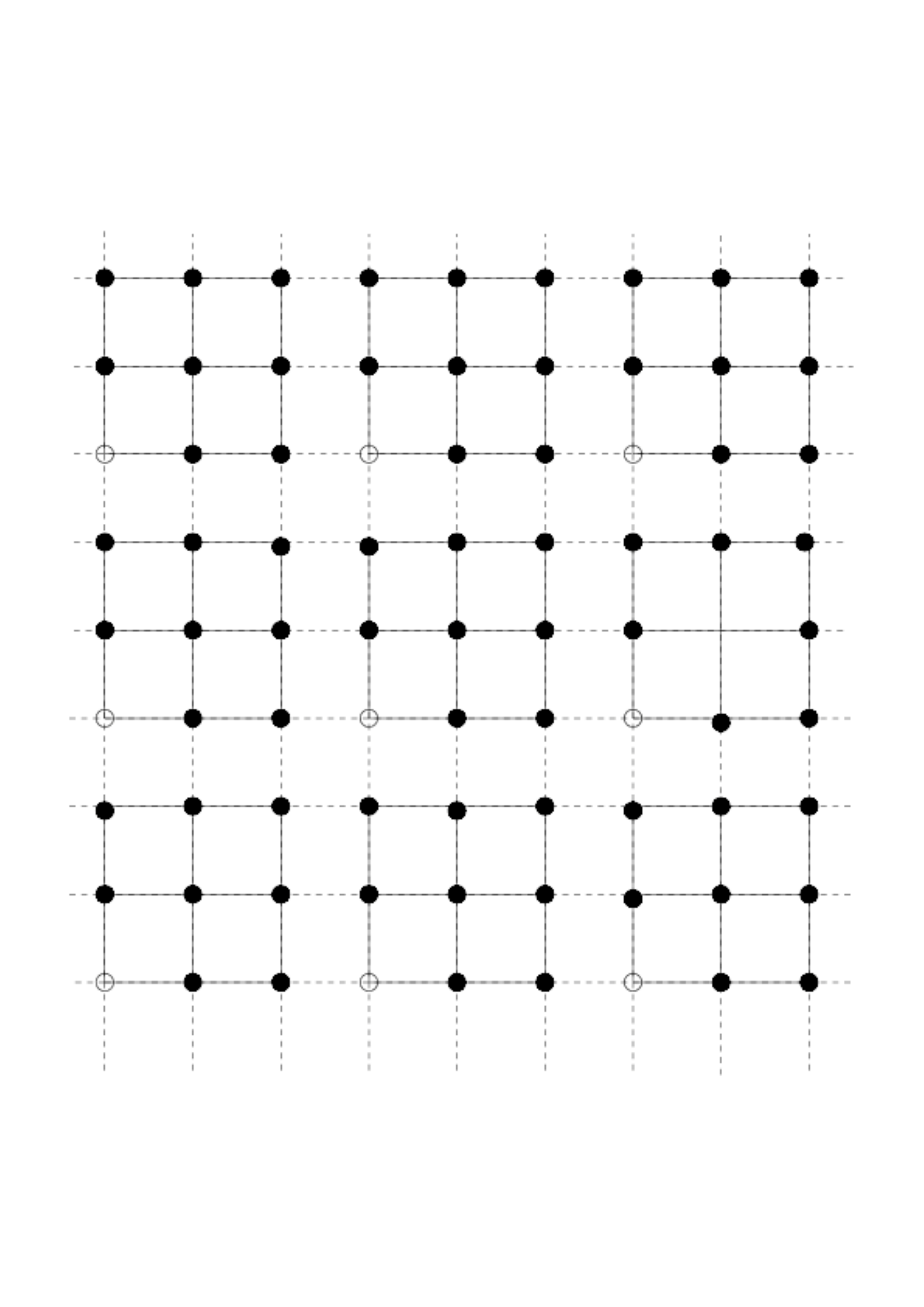}
   }
\caption{Examples of the three types of configurations considered in this paper are shown. In (a) a completely random configuration is shown in which $r \simeq 1$. In (b) and (c), the entire lattice is divided in periodic blocks of size $l \times l$, where $l=3$. The configuration in (b) is randomly periodic (RP) in that each block contains a single scatterer that is assigned to be a left rotator (open circle). The configuration in (c) is a completely periodic (CP) configuration, where the single left rotator in the same relative position in each block. In both (b) and (c) the initial ratio of rotators is $r=1/(l\times l-1)=1/8$.}\label{fig2}
\end{figure}

\section{Model}{\label{model}}
We start with a particle that moves along the bonds of a square lattice with constant unit speed (see figure \ref{fig1}(a)). When the particle arrives at a lattice site it encounters either a right or left rotator that rotates its velocity either to the right by an angle of $\theta=-\pi/2$ or left by an angle of $\theta=\pi/2$, respectively. Once the rotator scatters the particle it then \emph{flips} orientation either from right to left or from left to right. The particle's \emph{trajectory}, starting without loss in generality at the origin, is the sequence of positions $\{\mathbf{r}(t)\}_{t\geq 0}$ where $\mathbf{r}(0)=\mathbf{0}$ indicates the particle's \emph{initial position} and $\mathbf{r}(t)\in\mathbb{Z}^2$ indicates the particle's position at each integer $t\geq0$.

Here, we consider three different types of rotator configurations initially on the lattice. The first is a \emph{completely random} (CR) configuration in which the orientation of each rotator is chosen independently of the others, with the probability $(r+1)^{-1}\in[0,1]$ of being a right rotator. The \emph{ratio} $r=C_L/C_R$ is then the ratio of left to right scatterers, where $C_L$ is the concentration of left rotators and $C_R$ is the concentration of right rotators on the lattice. An example of a CR configuration is shown in Fig. \ref{fig2}(a) with ratio $r\simeq1$.

In the second type of configuration we consider, we first divide the lattice into periodic $\ell\times\ell$ blocks. In each block we initially let each rotator be a right rotator. We then randomly select $n\leq \ell^2$ rotators from each block and change their orientation so that they become left rotators. The resulting configuration of rotators is what we refer to as a \emph{rondom periodic} (RP) configuration. The reason for this designation is that we have enforced a periodic albeit random condition on the blocks that make up the configuration. It is worth mentioning that an important motivating fact for studying these configurations is that as $\ell\rightarrow\infty$ a random periodic configuration becomes a completely random configuration. An example of a RP configuration is shown in fig. \ref{fig2}(b).

In the third type of configuration we consider, we choose a single $\ell\times\ell$ block, again containing a single left rotator, which we periodically place over the entire lattice. The result is what we refer to as a \emph{completely periodic} (CP) configuration. A CP configuration has the property that it is invariant under vertical and horizontal translation by $\ell$ lattice sites since each block is identical. An example of a CP configuration is shown in fig. \ref{fig2}(c).

In both the RP and CP periodic configurations we think of the single left rotator, initially found in each block, as a \emph{defect} in the media through which the particle moves. In a RP configuration the defects appear randomly throughout the lattice except that there is exactly one in each $\ell\times\ell$ lattice block. In contrast, whereas in a CP lattice the defects are regularly spaced throughout the lattice and are in fact periodic (see figures \ref{fig2}(b) and (c)).

In general, a configuration of scatterers can be placed into two broad categories. The first is a \emph{random configuration} in which the orientation of the scatterers are generated via some random rule. The second is a \emph{deterministic configuration} where the orientation of each scatterer is uniquely determined by some fixed rule. For the initial configurations we consider, both CR and RP are random configurations, while CP configurations are deterministic. This fact will play key role in analyzing the dynamics of a particle moving through these different types of configurations.


\begin{figure}[htbp]
\begin{center}
\includegraphics[scale=0.35]{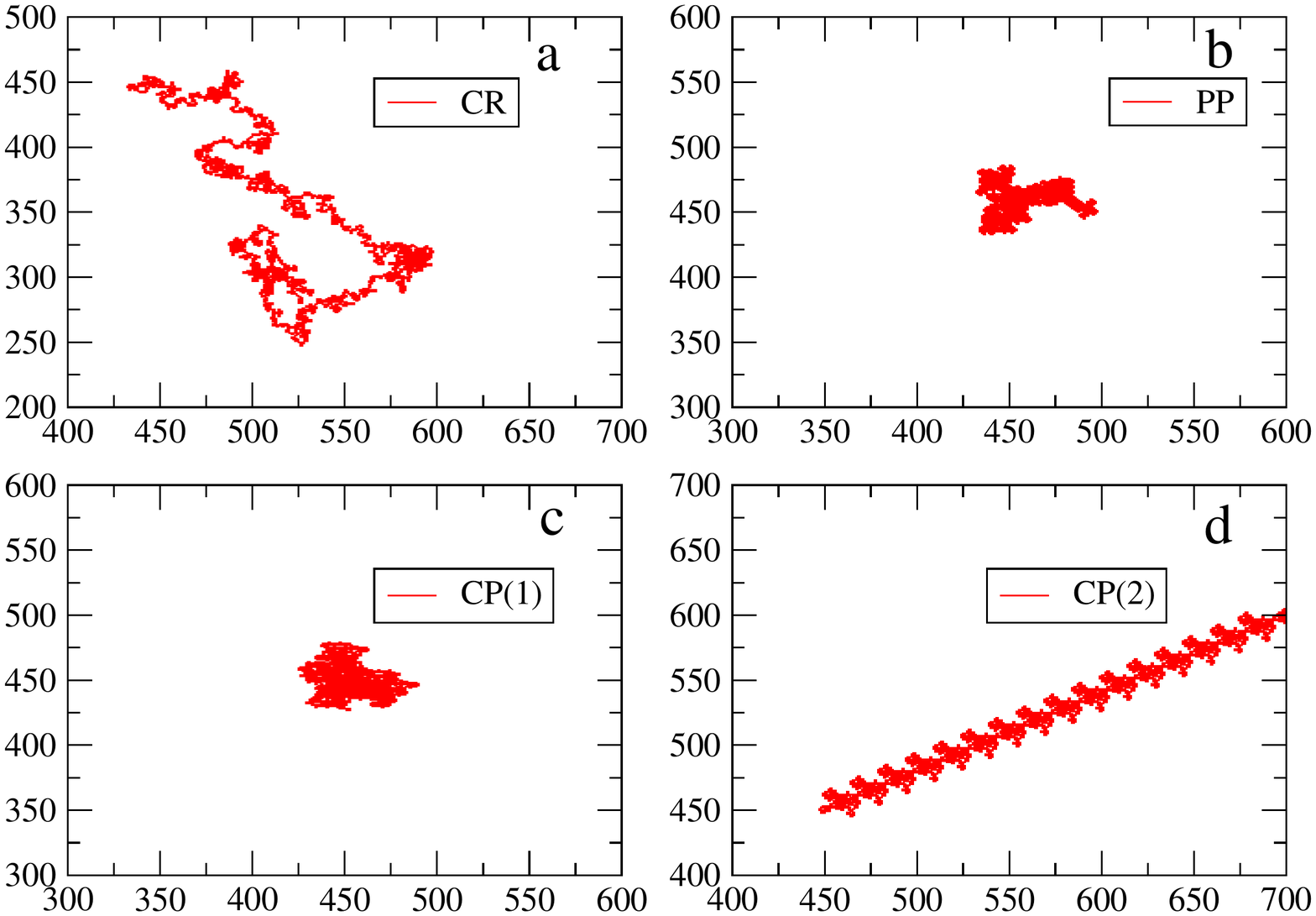}
\caption{In (a), a typical trajectory of the particle is shown for a completely random (CR) initial configuration with a ratio of $r=1$. In (b), a typical trajectory for a random periodic (RP) initial configuration is shown for $r=1/8$. In (c) and (d), a non-propagating and propagating trajectory are shown, respectively, for two different completely periodic (CP) initial configurations. These are labeled (CP1) and (CP2).}
\label{fig3}
\end{center}
\end{figure}

Our goal in studying this model is to determine to what extent the particle's dynamics is influenced by the amount of periodicity, i.e. \emph{order}, present in each of the configuration types CR, RP, and CP for various values of $r\in[0,1]$. To carry out this analysis, we characterize the trajectory of the particle by calculating the following four quantities (1)-(4).\\
\textbf{(1)} The \emph{mean square displacement} (MSD) of particle at time t: The particle's MSD is defined by
\begin{equation}
\Delta(t) = \langle|\mathbf{r}(t)-\mathbf{r}(0)|^2\rangle \ \text{for} \ t\geq 0,
\end{equation}
where $|\cdot|$ is the Euclidian distance in the plane and $\langle\cdot\rangle$ is the average over many initial configurations of a certain type, e.g. CR and RP for a fixed value of $r$. This average is computed by first generating $N$ such initial configurations. Allowing the particle to move through each configuration similarly generates $N$ trajectories $\{\mathbf{r}_i(t)\}_{t\geq 0}$ for $1\leq i\leq N$. The MSD of the particle is formally defined to be the limit 
\begin{equation}\label{eq:3}
\Delta(t)=\lim_{N\rightarrow\infty}\frac{1}{N}\sum_{i=1}^N|\mathbf{r}_i(t)-r_i(0)|^2 \ \ \text{for} \ \ t\geq 0.
\end{equation}
if this limit exists. In practice, the particle's MSD is found numerically by taking $N$ to be large enough that $\Delta(t)$ appears to converge to a specific number each $t\leq T$, where the number $T$ is the largest time we wish to compute.\\
\textbf{(2)} The nature of the boundary of visited sites: This is done by finding the fractal dimension $d_f$  of boundary of visited sites, which can be calculated from the average \emph{area} $\langle A(t)\rangle$ of lattice sites visited by time $t\geq 0$ and the average \emph{boundary} $\langle B(t)\rangle$ of the lattice sites visited by time $t\geq 0$. Formally, for a given initial configuration of rotators $A(t)$ is the number of distinct lattice sites visited by the particle by time $t$ and $B(t)$ is the number of boundary sites of this region. Here, $\langle\cdot\rangle$ denotes the average over many realizations of a specific type of configuration and is defined similar to equation \eqref{eq:3}. We define a site to be a \emph{boundary point} of this region if any of its nearest neighbors is not visited by the particle by time $t$. For a smooth boundary, $\langle B(t)\rangle \propto \langle A(t)\rangle^{1/2}$ and the boundary has dimension 1. With this in mind, the \emph{fractal dimension} of a boundary is defined to be the number $d_f\geq0$ such that $\langle B(t)\rangle \propto \langle A(t)\rangle^{d_f/2}$, if this number exists.\\
\textbf{(3)} The \emph{time-dependent ratio of right to left rotators} $r(t)= C_L(t)/C_R(t)$: This is defined to be the ratio of right to left scatterers on the lattice sites visited by the particle by time $t\geq 0$. Here, $C_R(t)$ and $C_L(t)$ are the concentration of right and left rotators, respectively, on the set of lattice sites visited by the particle by time $t\geq 0$.\\
\textbf{(4)} The rotator-rotator orientation correlation $F_{in}(t)$ and the time-dependent ratio of right to left rotators $r_{in}(t)$ inside the visited sites: The \emph{rotator-rotator orientation correlation}  is defined by
\begin{equation}
F_{in}(t) = \langle O(i,t) O(i+1,t)\rangle \ \text{for} \ t\geq0,
\end{equation}
where $O(i,t)$ is the orientation of the rotator at site $i$ at time $t$. The average $\langle\cdot\rangle$ is taken over all lattices sites inside the region the particle visits by time $t$. The rotator-rotator orientation correlation is calculated in the horizontal direction only. The reason is that the system we consider is isotropic, so that the same results are obtained as well for the vertical direction.

The ratio $r_{in}(t)$ is defined by
\begin{equation}
r_{in}(t)=\frac{C_{in R}(t)}{C_{in L}(t)} \ \text{for} \ t\geq 0,
\end{equation}
where $C_{in R}(t)$ and $C_{in L}(t)$ are the concentration of right and left rotators, respectively, inside the lattice sites visited by the particle by time $t$.

\begin{figure}[htbp]
\begin{center}
 \subfigure[]{
  \includegraphics[scale=0.3]{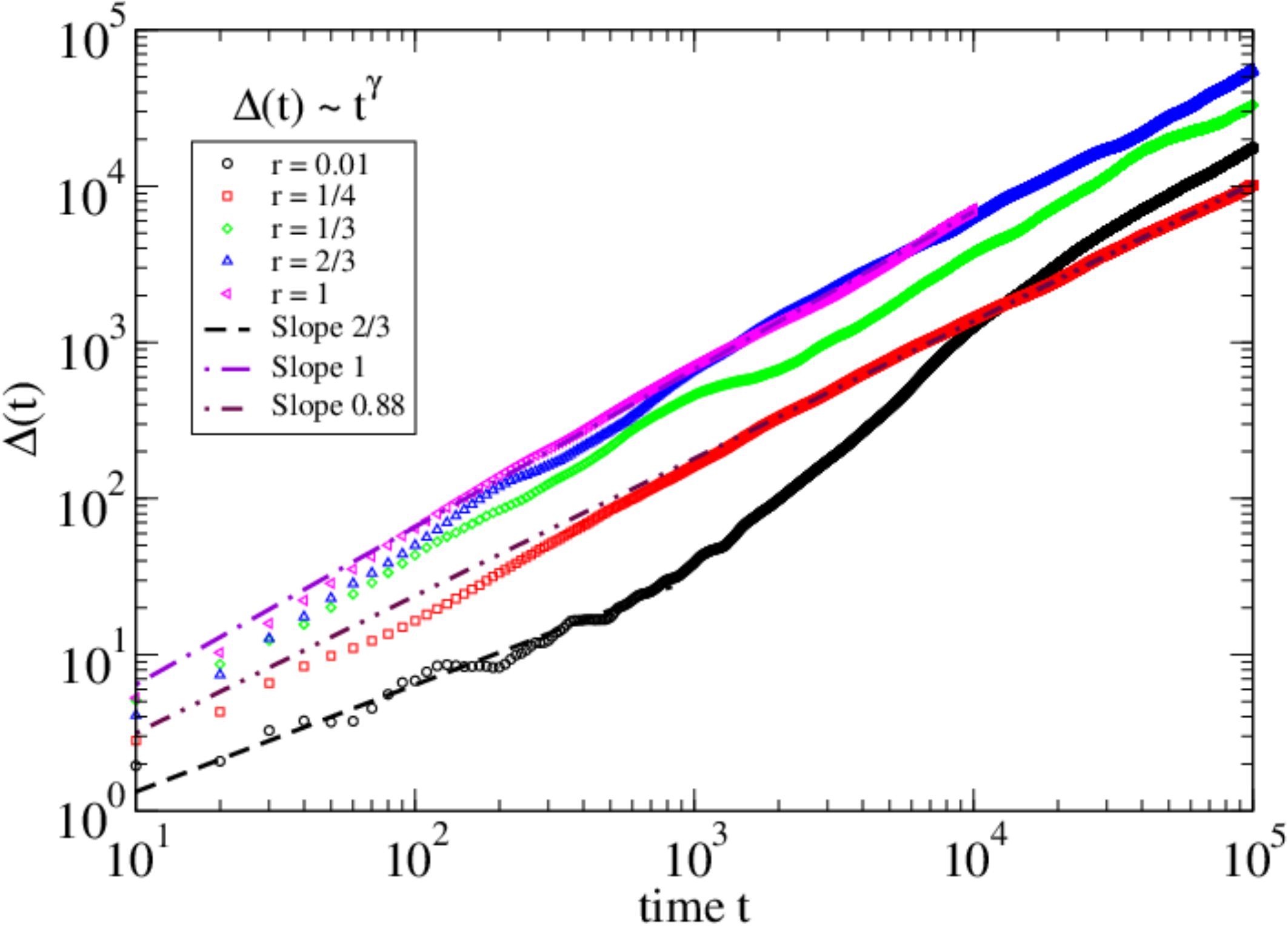}
   }
 \subfigure[]{
  \includegraphics[scale=0.25]{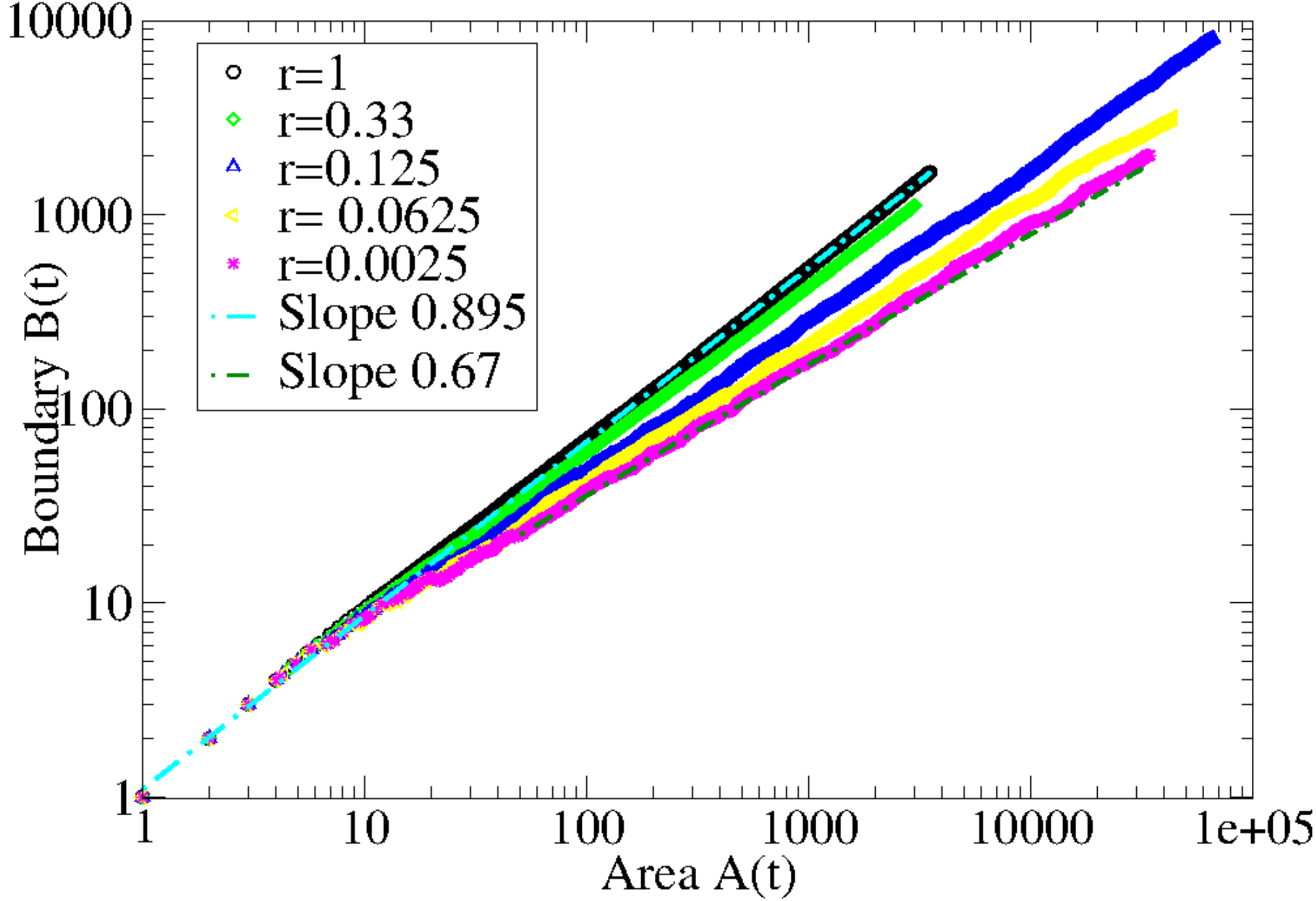}
   }

\caption{(color online)
(a) A plot of the particle's MSD, $\Delta(t)$ for $0\leq t\leq 10^5$ for completely random (CR) initial configurations for a number of different ratios $r\in[0,1]$ of right to left rotators. For $r\simeq 0$, the initial configuration of scatterers is nearly homogenous, whereas for $r\simeq 1$ the number of right to left rotators on the lattice is nearly identical. For these $r$-values, $\Delta(t) \propto t^{\alpha}$ were $\alpha $ varies from $\alpha\simeq 2/3$ to $\alpha\simeq 1$ as $r$ increases from $r\simeq 0$ to $r\simeq 1$. Hence, as $r$ increases there is a transition in the particle's dynamics from subdiffusion to anomalous diffusion. For $r=0$, since we are approaching homogeneous
configuration hence at late time particle might start propagating hence dynamics is faster and $r \simeq0$ curve crosses with other $r$ values. (b) A plot of the boundary $B(t)$ vs. the area $A(t)$ for a number of $r$-values is shown. Here, $B(t) \simeq A(t)^{\alpha_1}$ with $\alpha_1 \ne 1/2$ implying that the boundary is fractal with fractal dimension $d_f = 2 \alpha_1$. The fractal dimension ranges from $1.34\pm(0.02) \simeq 4/3$ for $r\simeq 0$ to $1.794\pm(0.03)\simeq 7/4$ for $r\simeq 1$. This suggests that the nature of the boundary changes from that of a self-avoiding walk to that of a percolating cluster in 2-d as $r$ increases from nearly 0 to 1.}
\label{fig4}
\end{center}
\end{figure}

\section{Results}\label{results}
In this section we discuss the dynamics and asymptotic state of particle moving over the three distinct types of initial configurations described in section \ref{model}, using quantities \textbf{(1)}-\textbf{(4)} also defined in section \ref{model}.

\subsection{Complete Random (CR)}\label{cr}
Here, the particle starts on a lattice with a completely random initial configuration of right and left rotators for a fixed ratio $r$. We note that, for a single realization of a CR configuration the particle's trajectory is completely deterministic. However, when averaged over many such configurations, the particle motion acquires the probabilistic interpretation of moving to the right with probability $(r+1)^{-1}$ each time it encounters a new lattice site. A typical trajectory of the particle over a specific realization of a CR configuration is shown in fig. \ref{fig3}(a), which is very similar to a 2-d random walk.

When $r\simeq 0$, the majority of the lattice's rotators are initially oriented to the right, so that the configuration is nearly homogeneous. When $r\simeq 1$, there are initially about the same number of right rotators as there are left rotators on the lattice. For $r$-values ranging from $r\simeq 0$ to $r=1$, we investigate the particle's motion in the CR case by calculating each of the quantities \textbf{(1)}-\textbf{(4)} described in section \ref{model}.

(1) As we increase the ratio $r$, the particle's MSD given by $\Delta(t)\propto t^{\alpha}$ changes continuously from $\alpha\simeq 2/3$ for $r\simeq 0$ to $\alpha\simeq 1$ for $r\simeq 1$. Hence, there is a continuous transition from subdiffusion with an exponent of $\alpha =2/3$ to anomalous
diffusion with an exponent of $\alpha =1$ as $r$ increases from nearly 0 to 1 (see fig. \ref{fig4}(a)). For $r=1$, the reason the particle's diffusion is anomalous is that the probability distribution of its position is non-Gaussian.

(2) The boundary of the sites visited by the particle behaves like $B(t) \propto A(t)^{d_f/2}$ where $d_f$ varies from $d_f=1.34 \simeq 4/3$ for $r\simeq 0$, the fractal dimension of a self-avoiding walk \cite{saw}, to $d_f=1.794$ for $r\simeq 1$, the fractal dimension of a percolation cluster in 2-d \cite{pc}. In Fig. \ref{fig4}(b) a number of plots of $B(t)$ vs. $A(t)$ for different values of $r$ are shown.

\begin{figure}[htbp]
\centering
 \subfigure[]{
  \includegraphics[height=1.5in,width=2.4in]{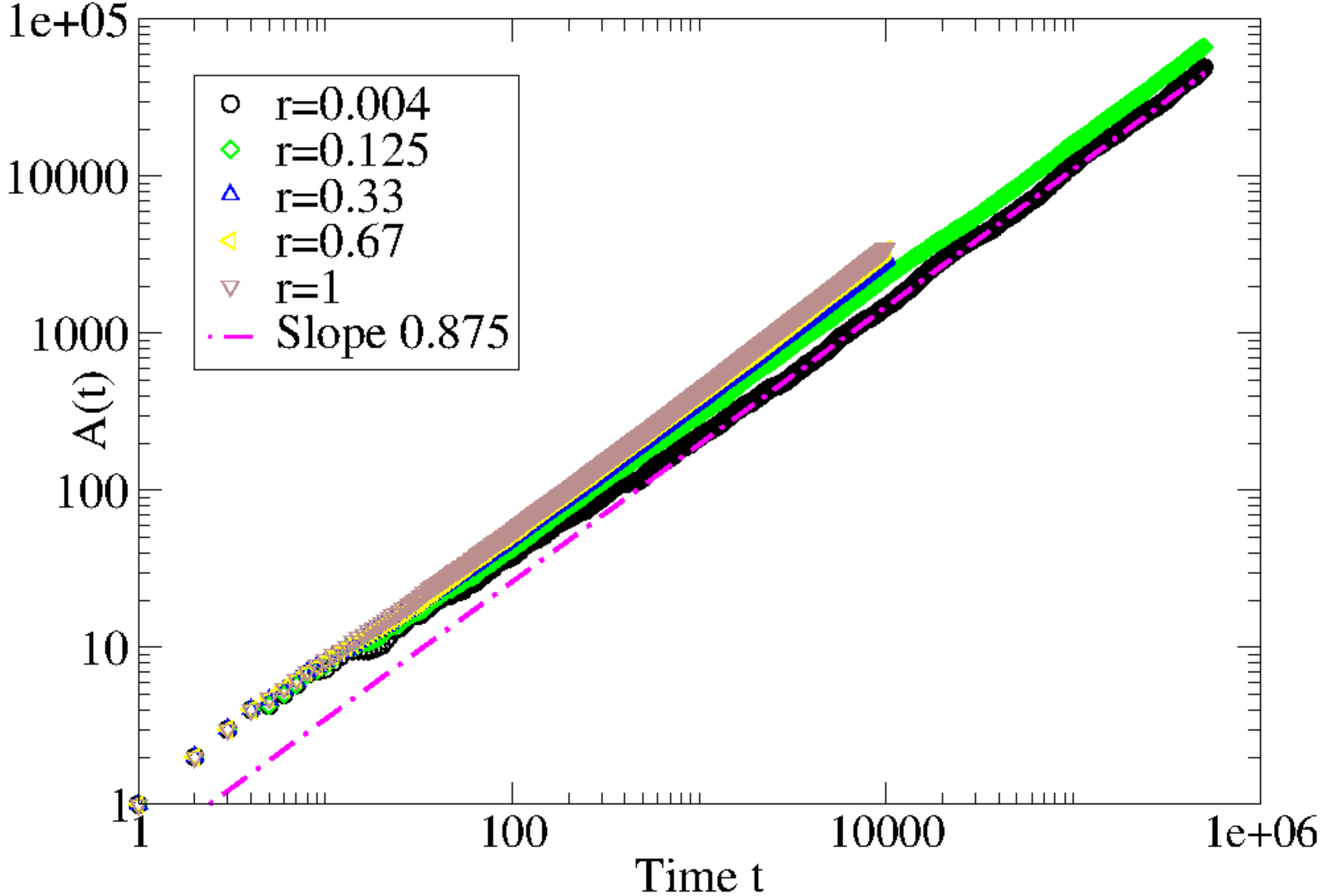}
   }
   \subfigure[]{
  \includegraphics[height=1.5in,width=2.4in]{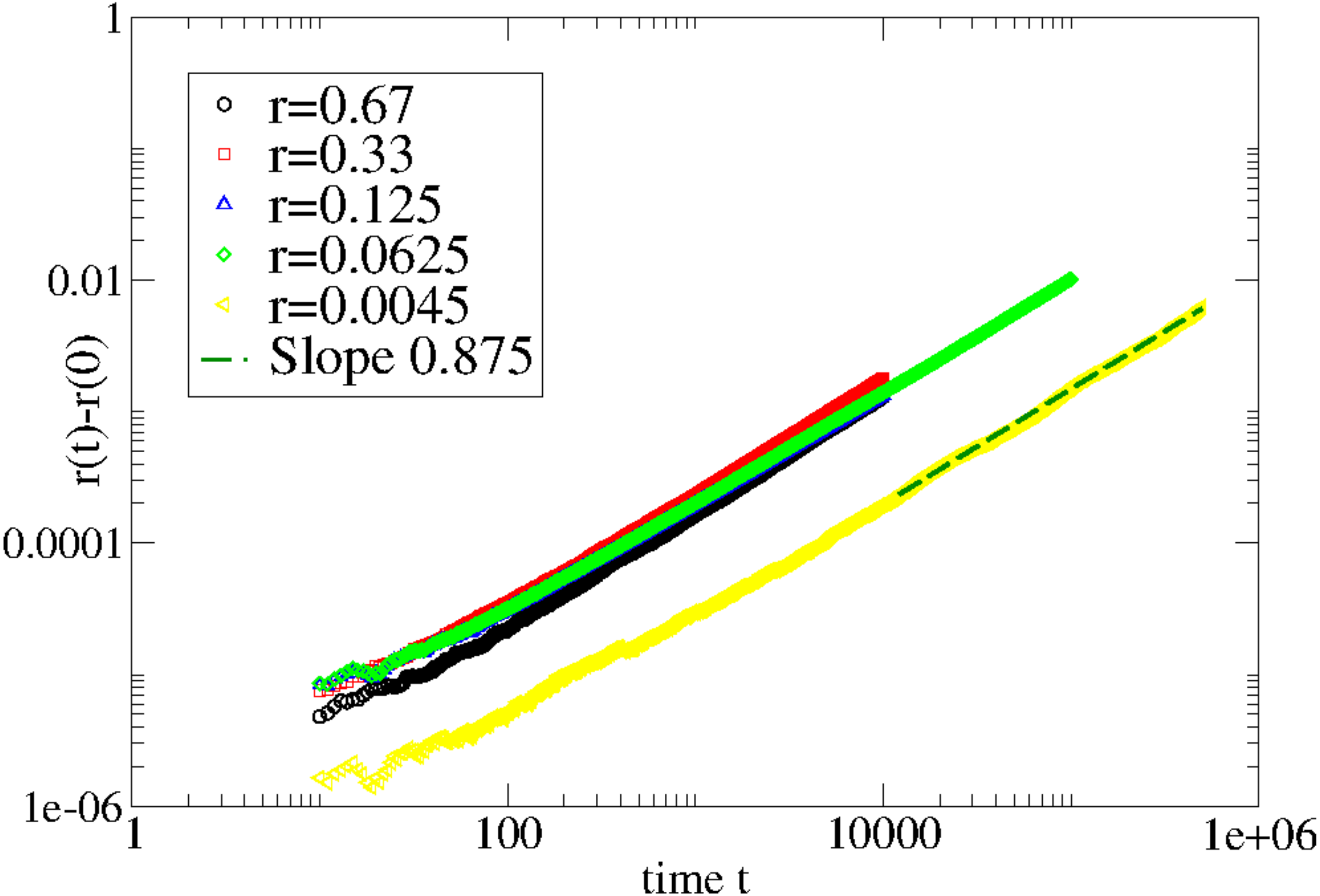}
   }
\caption{(color online) (a) A plot of the area $A(t)$ the particle visits while moving over a CR configuration for different values of the ratio $r$. (b) The difference of the ratios $r(t)-r(0)$ is also shown for various $r$-values of the particle moving over a CR configuration. The area $A(t)$ and difference $r(t)-r(0)$ increase with the unique exponent $\gamma \simeq 0.875 = 7/8$ (dashed line).}
\label{fig5}
\end{figure}

(3) As $r$ increases, the area $A(t)$ and the difference of ratios $r(t)-r(0)$ on those sites the particle has visited also increases, both with the same power $\gamma \simeq 0.875 = 7/8$. This is shown in figures \ref{fig5}(a) and (b), respectively. We note that the area $A(t)$ is increasing because the particle is exploring more and more of the lattice as time increases. The difference $r(t)-r(0)$ increases because, as particle is moving on the lattice, it tries to make ratio inside the visited site homogeneous. Hence, the ratio $r(t)$ inside the sites visited by the particle approaches 1, but at the same time the particle is exploring more and more lattice sites. 

(4) We find that the rotator correlation function $F_{in}(t)$ approaches 0 while the ratio $r_{in}(t)$ approaches 1 as $t$ increases, for the set of $r$-values we consider (data not shown). This means that, while moving on the lattice the  particle tries to randomize the orientation of the rotators it visits while at the same time leaving behind the same number of right and left rotators.


\subsection{Random Periodic (RP)}\label{pp}
Now we study the motion of particle when the initial configuration of rotators is initially arranged in random periodic manner, as is shown in fig. \ref{fig2}(b). In a random periodic configuration the lattice is first broken into disjoint $\ell\times \ell$ blocks, which contains all right rotators. From each of these blocks we randomly select $n\leq \ell^2$ rotators and switch the orientation of each from from right to left. We call the resulting configuration of rotators a \emph{random periodic} configuration with ratio $r=n/(\ell^2-n)$. From a physical point of view,  left rotators in each lattice block can be thought of as \emph{defects} in the lattice configuration or medium through which the particle moves. A RP configuration is essentially a CR configuration in which we have stipulated that not only do we globally have a specific ratio $r$ but also in each $\ell\times\ell$ block of the lattice has this ratio. Hence, as $\ell\rightarrow\infty$ a RP configuration becomes a CR configuration for some fixed value of $r$.

\begin{figure*}[htbp]
\centering
\subfigure[]{
        \includegraphics[height=1.8in,width=3.0in]{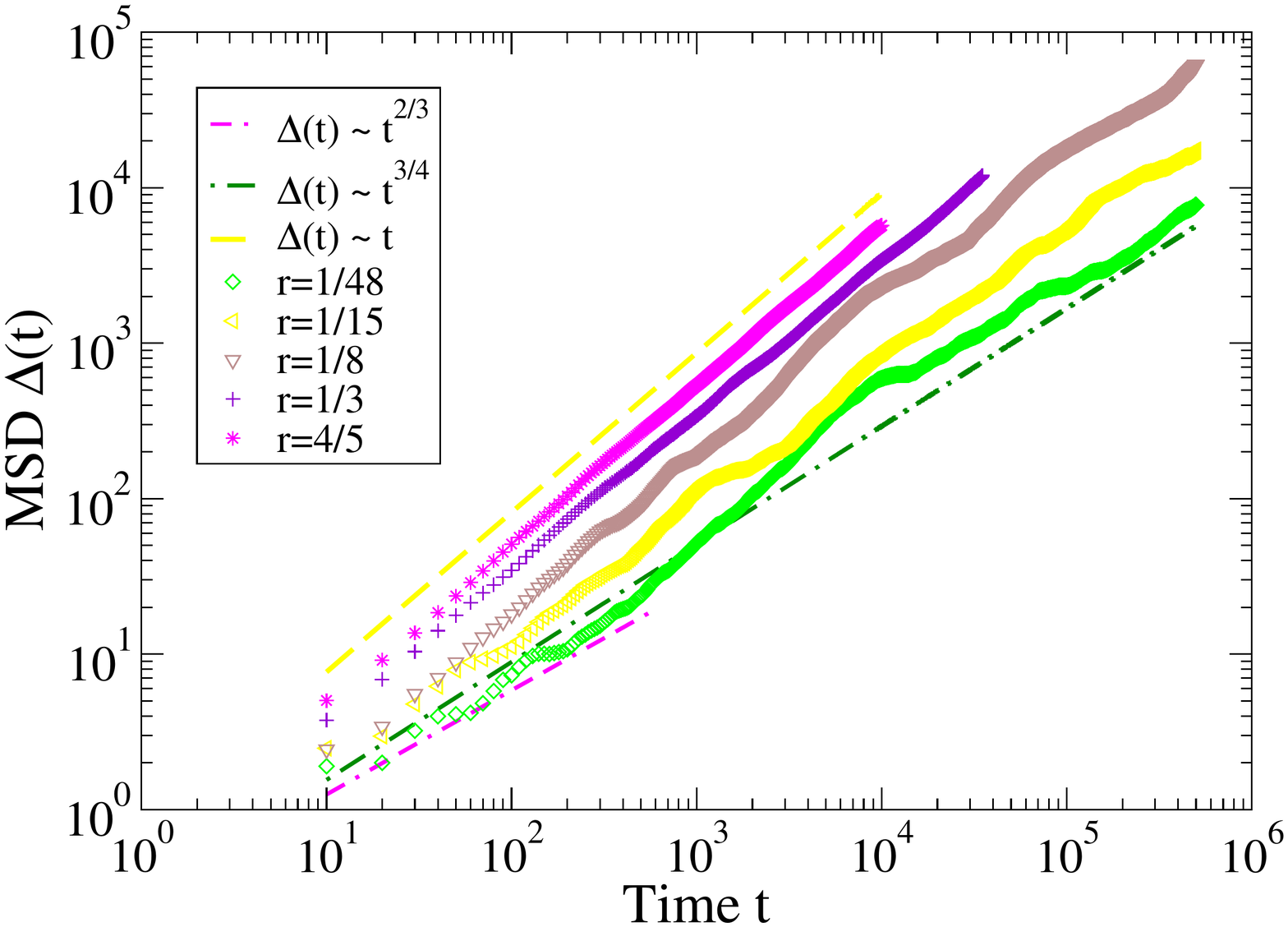}
         }
\subfigure[]{
        \includegraphics[height=1.5in,width=2.4in]{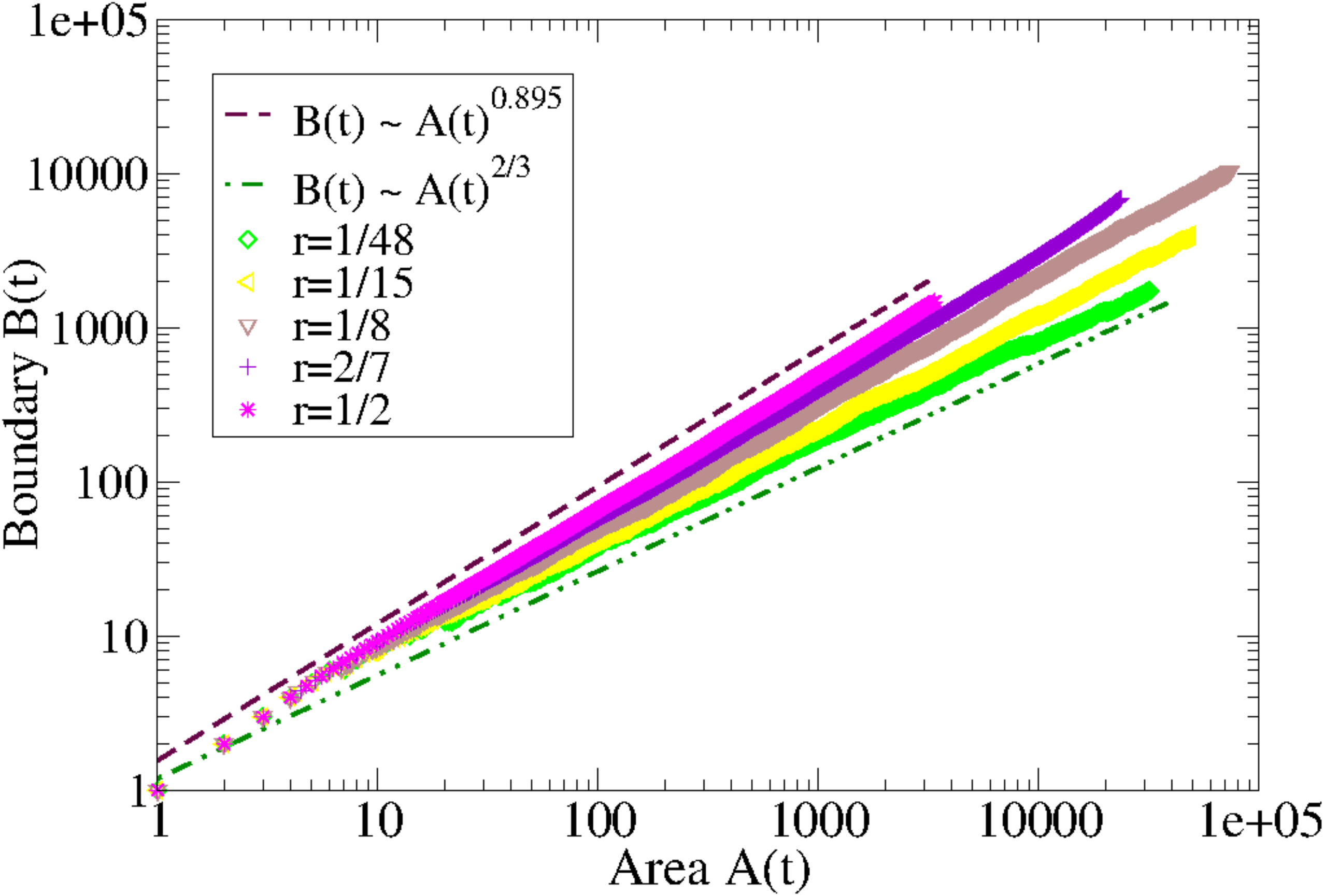}
        }
\subfigure[]{
        \includegraphics[height=1.5in,width=2.4in]{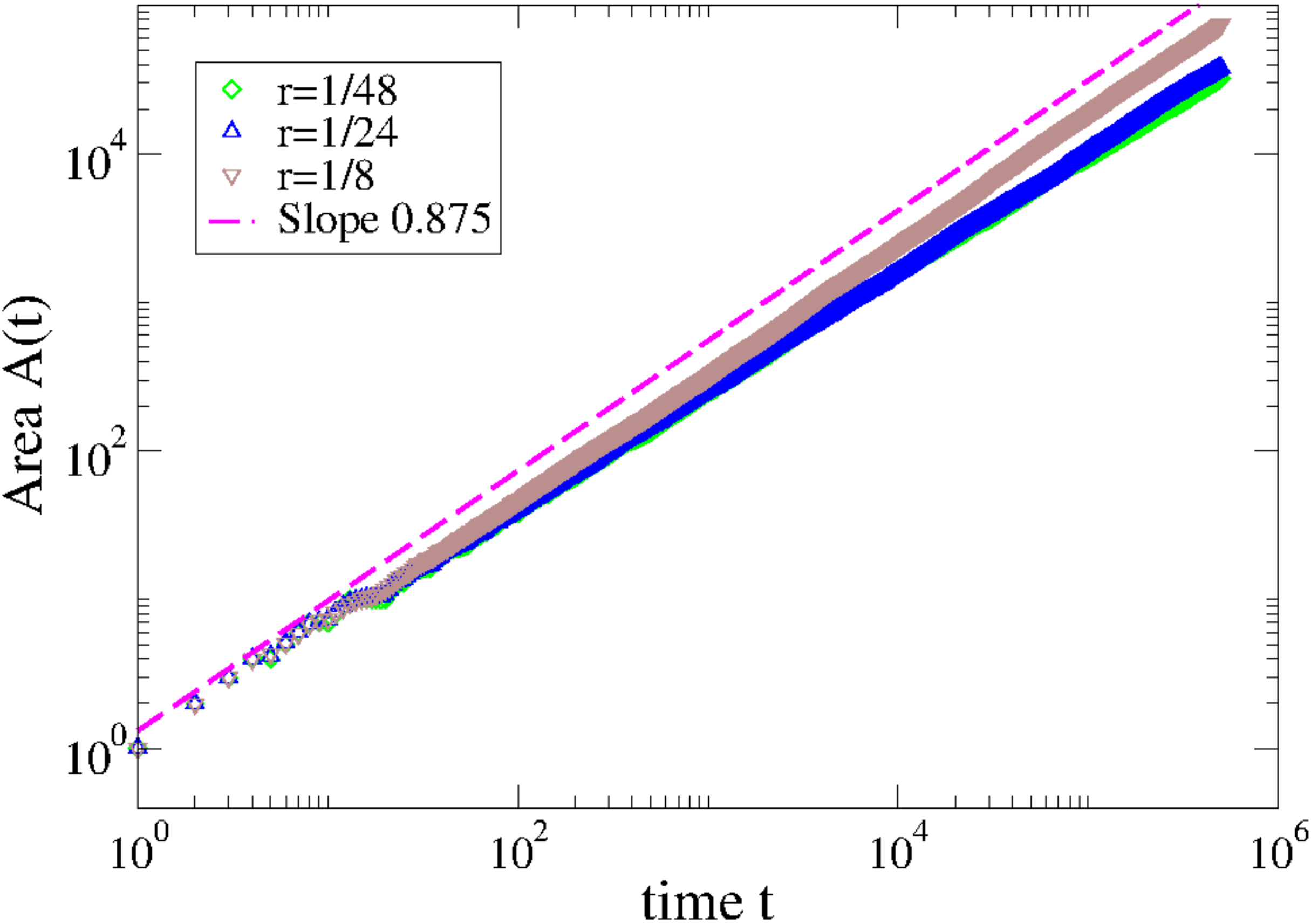}
        }
\subfigure[]{
        \includegraphics[height=1.5in,width=2.4in]{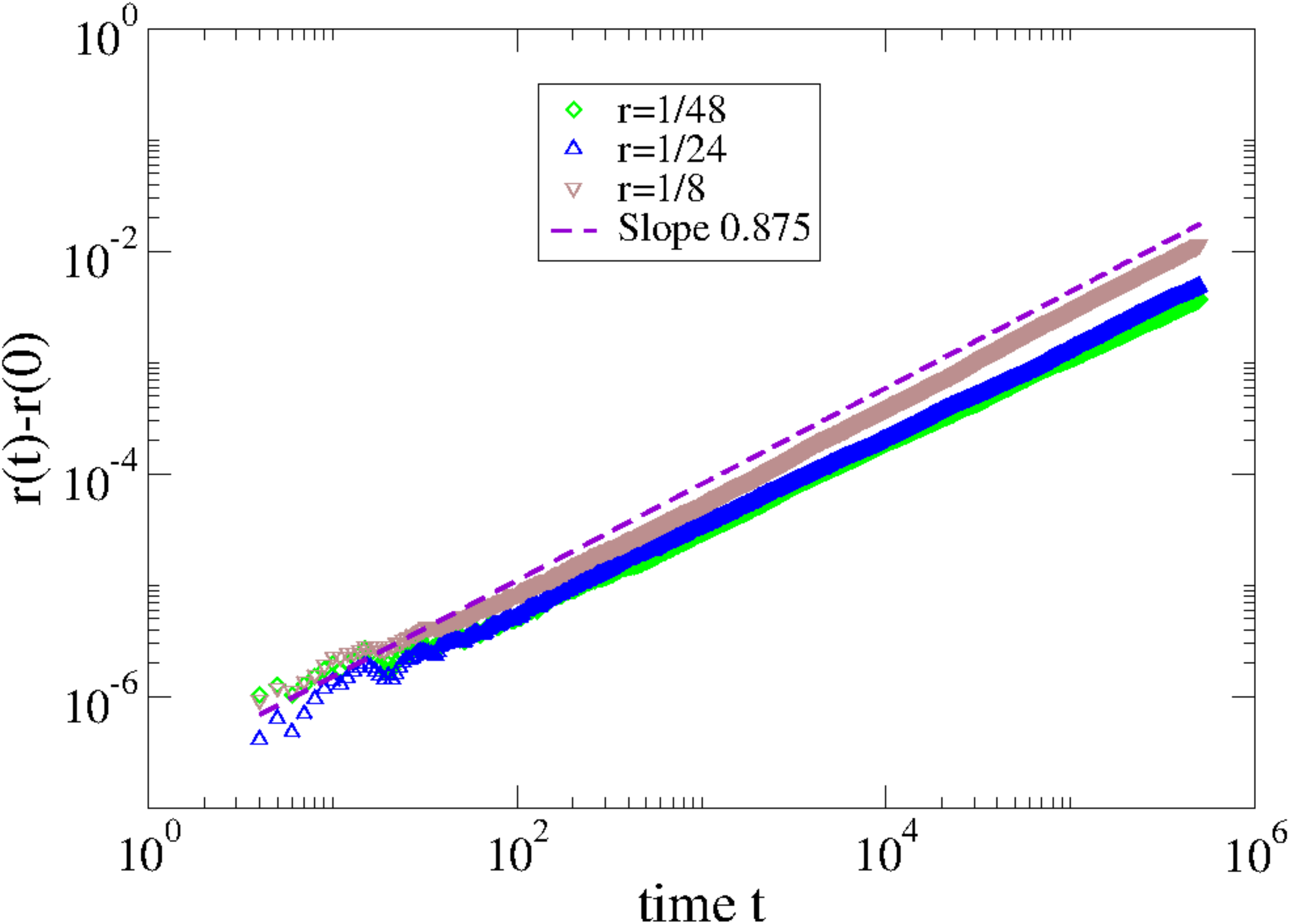}
         }
\caption{(color online) For a number of different $r$-values, a plot of (a) the particle's MSD, (b) the fractal dimension $d_f$ of its trajectory, (c) the area $A(t)$ of sites visited by the particle, and (d) the time-dependent difference $r(t)-r(0)$ are plotted. In each case the initial configuration of rotators is randomly periodic (RP).}
\label{fig6}
\end{figure*}

We find that the dynamics of a particle moving on randomly periodic initial configuration
is statistically identical to the particle's motion over a CR configuration (see section \ref{cr}). Similar to computing the MSD of a CR configuration for a specific ratio $r$, we calculate the MSD for a RP configuration by averaging over a large number RP configurations for a fixed $\ell\in\mathbb{Z}$ and $r\in[0,1]$. In figure \ref{fig6}(a) we plot the particle's MSD for various ratios of $r$. Again we find that $\Delta(t) \propto t^{\beta}$, where $\beta\simeq 2/3$ for $r\simeq 0$, which increases until $\beta\simeq 1$ for $r\simeq 1$. That is, the particle's motion transitions from being subdiffusive to exhibiting anomalous diffusion as $r$ increases.

In  fig. \ref{fig6}(b) we plot the boundary of visited sites $B(t)$ vs. the area of visited sites $A(t)$ for various ratios of $r$. Again we find that $\langle B(t)\rangle \simeq \langle A(t)\rangle^{d_f/2}$, where the fractal dimension $d_f$ varies from  $1.34$ to $1.794$ as $r$ increases from nearly 0 to 1. Hence, the particle moving over a RP configuration has a trajectory that is statistically similar to a self-avoiding walk \cite{saw} for small $r$ and is similar to a percolation cluster type in 2-d for $r$ near 1 \cite{pc}.

We also calculate the time-dependent area of visited sites $A(t)$ and the difference of ratios $r(t)-r(0)$. Again we find that both vary with the same exponent $\gamma \sim 0.875 = 7/8$ for different ratios of $r$, similar to the case in which the configuration is CR (see figures \ref{fig6}(c) and (d), respectively).


\begin{figure}[htbp]
\begin{center}
\includegraphics[scale=0.25]{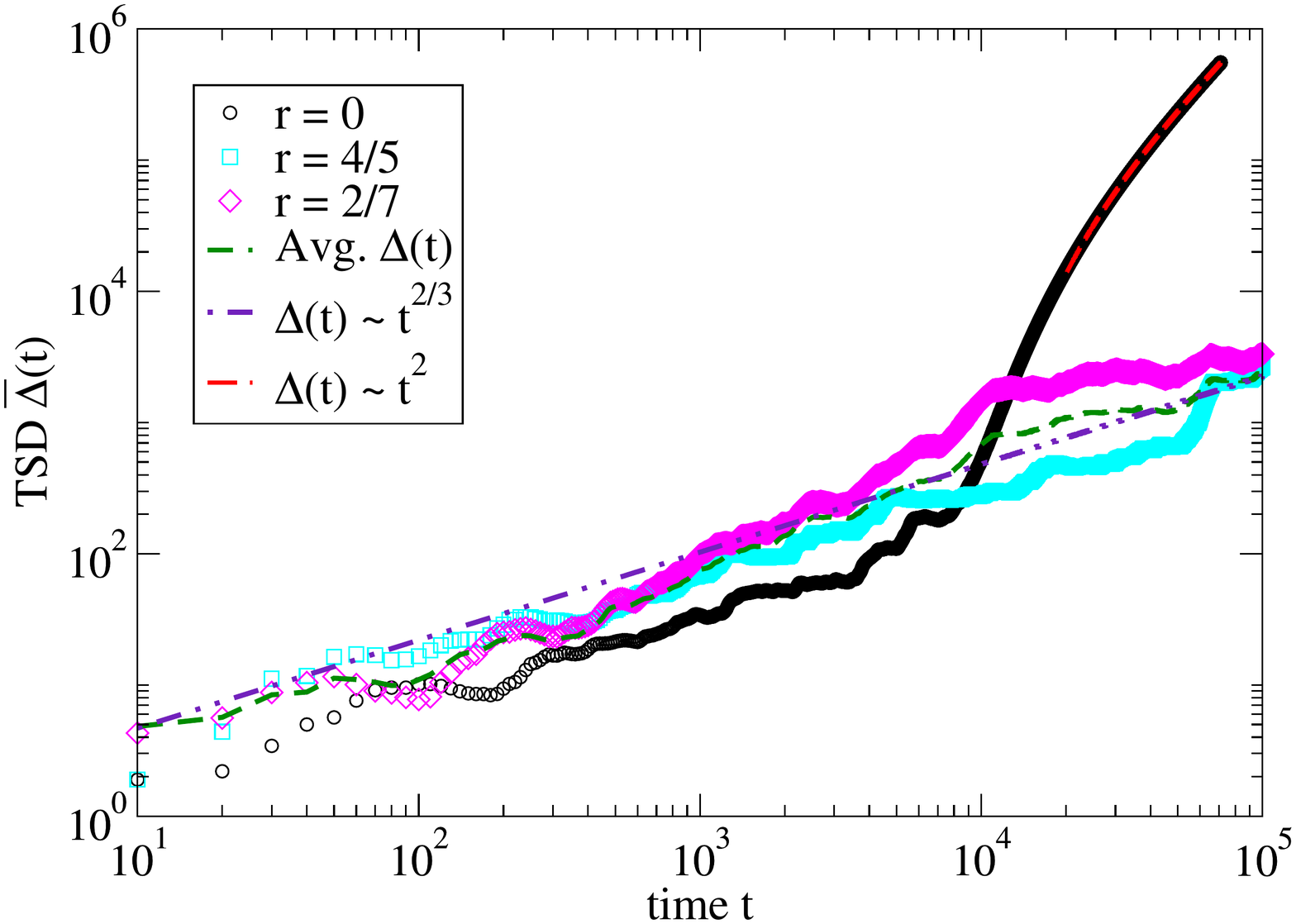}
\caption{(color online) The time-averaged square displacement (TSD) $\bar \Delta(t)$ of the particle is shown for  completely periodic (CP) configurations with ratios $r=0$, $2/7$ and $4/5$. These periodic configurations are created by making blocks of size $3 \times 3$ and in 
each block we label the rotators from $1,2,3....9$ as we go from left to right. For ratio $r=0$ all the rotators in each block is of right rotators. For $r=2/7$, rotators at position 1 and 2 are left and all others are right rotators. For $r=4/5$, rotators at position 1,2,4 and 5 are 
left and all others are right. But for each ratio above configurations are not unique and one can create many other completely periodic
configurations with same ratio. For $r=0$ the particle begins propagation after a transient time of $t=9977$. In contrast, for $r=2/7$ and $r=4/5$ the particle's trajectory is non-propagating. For all three ratios the initial subdiffusion behavior is same where the TSD, $\bar \Delta(t) \simeq t^{2/3}$. For comparison, the purple dot-dashed line has slope 2/3 and the green dot-dashed line give the average behavior of the two ratios $r=2/7$ and $r=4/5$. The red dashed line has a slope of 2 to show the propagation of the particle after its transient subdiffusive behavior for $r=0$.}
\label{fig7}
\end{center}
\end{figure}

\subsection{Complete Periodic (CP)}\label{cp}
We now study the dynamics of particle moving on a completely periodic configuration of rotators. As in a RP configuration, the lattice is first partitioned into a number of $\ell\times\ell$ blocks. The blocks we choose are the same as the blocks we use to generate a RP configuration, i.e. $\ell\times\ell-1$ rotators are right rotators so that again $r=n/(\ell^2-n)$. The difference is that in a completely periodic (CP) configuration each block is identical, not randomly generated. Because of this, a CP configuration is invariant under vertical and horizontal translation by $\ell$ lattice sites. An example of a CP configuration is shown in fig. \ref{fig2} (c), in which $n=1$, $\ell=3$, and $r=1/8$. As in the RP case, we think of the single left rotator in each lattice block as a defect in the media in which the particle moves.

Completely periodic configurations differ from CR and RP configurations in at least one significant way. While CR and RP configurations are randomly generated, CP configurations formally are not. That is, although there are infinitely many CP configurations we consider each on of them individually. One consequence of this is that the particle's MSD over a CP configuration is simply its square displacement $\Delta(t)=|\mathbf{r}(t)-\mathbf(0)|$, which can behave very erratically (see \cite{meng1994}, for instance). To more easily analyze the particle's displacement over a CP configuration we consider its time-averaged square displacement given by
\begin{equation}
\bar\Delta (t) = \frac{1}{t}\sum_{\tau = 0}^{t}|\mathbf{r}(\tau)-\mathbf{r}(0)| \ \text{for} \ t>0.
\label{eqtmsd}
\end{equation}
A number of properties regarding a particle's TSD can be found in \cite{ben2014}. Importantly, if a particle exhibits subdiffusion, diffusion, superdiffusion, or propagation based on its MSD, then the same holds with respect to its TSD.

A number of CP configurations have been studied by Cohen and Meng in \cite{meng1994}. For these configurations the authors found that, after some initial transient time the particle typically starts propagating in one direction through the lattice. However, for many CP configurations the initial transient time before propagation is very large and difficult to determine numerically.

Here, we study a much larger class of CP configurations. Our goal is to understand how the particle's motion differs as it moves over a RP configuration verses a CP configuration. Another is to determine whether there is any similarity in the particle's dynamics for the case in which the particle starts to propagate in a computationally feasible amount of time verses when it does not. We call the first of these types of trajectories \emph{propagating trajectories} and the latter \emph{non-propagating trajectories}, respectively.

We find that the dynamics of the particle before it starts to propagate in both the propagating and non-propagating case are statistically identical. This we show based on the particle's TSD, the fractal dimension $d_f$ of its trajectory, and the rotator-rotator correlation function $F_{in}(t)$ and the ratio $r_{in}(t)$ for a fixed initial ratio $r$. For these three characteristics we find the following.

We find that in those trajectories that propagate there are two distinct stages in the particle's dynamics. In the first, the particle subdiffuses with a TSD of $\bar \Delta(t) \propto t^{2/3}$ for some finite amount of time depending on the particular CP configuration under consideration. In the second stage the particle abruptly transitions to a ballistic trajectory where $\bar{\Delta(t)} \propto t^{2}$. Here, the propagation is always in a single direction through the lattice. We call these two stages the particle's transient \emph{preparation stage} and its \emph{propagation stage}, respectively.

\begin{figure}[htbp]
\begin{center}
\centering
\subfigure[]{
        \includegraphics[height=1.5in,width=2.4in]{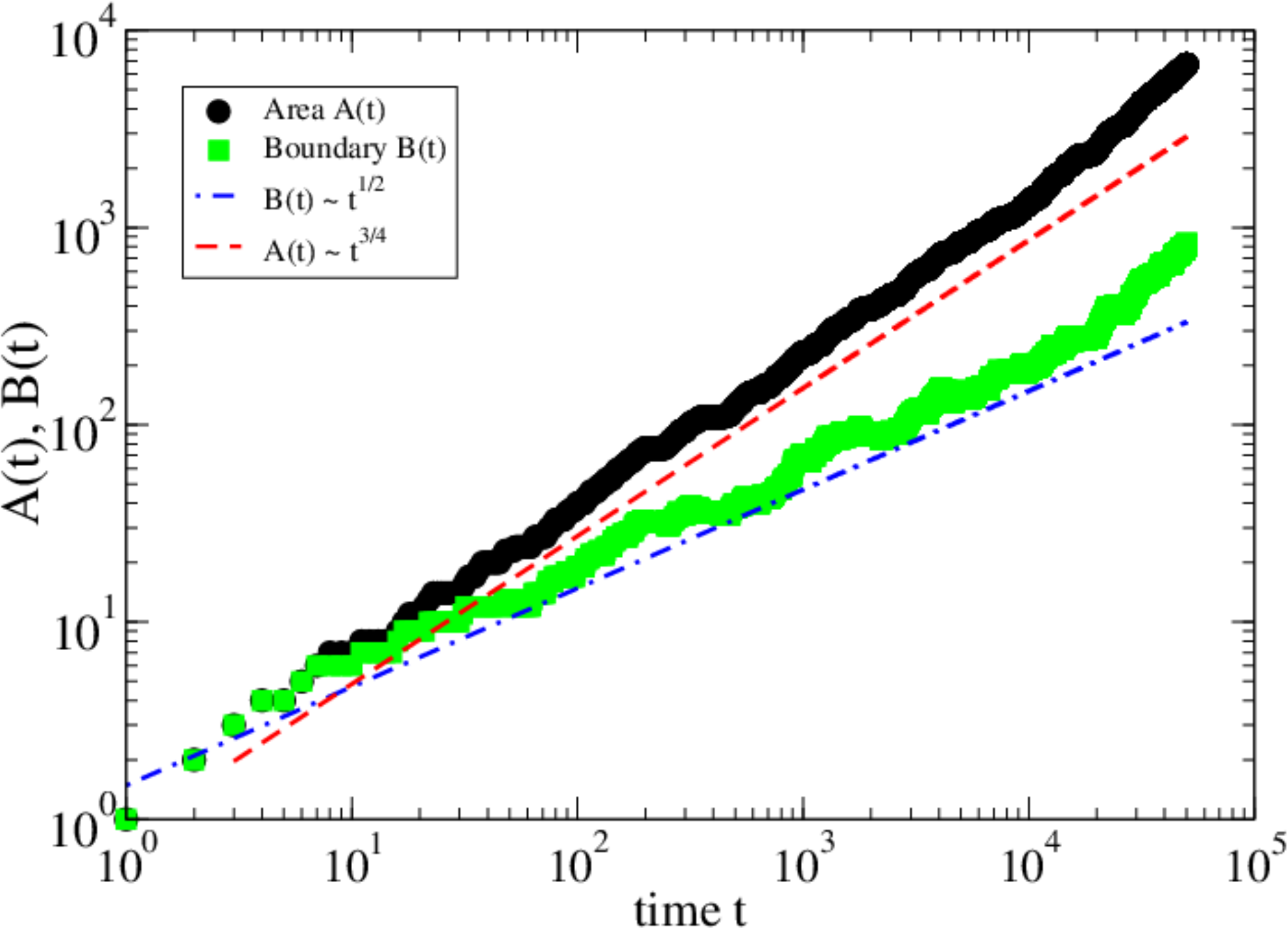}
         }
\subfigure[]{
        \includegraphics[height=1.5in,width=2.4in]{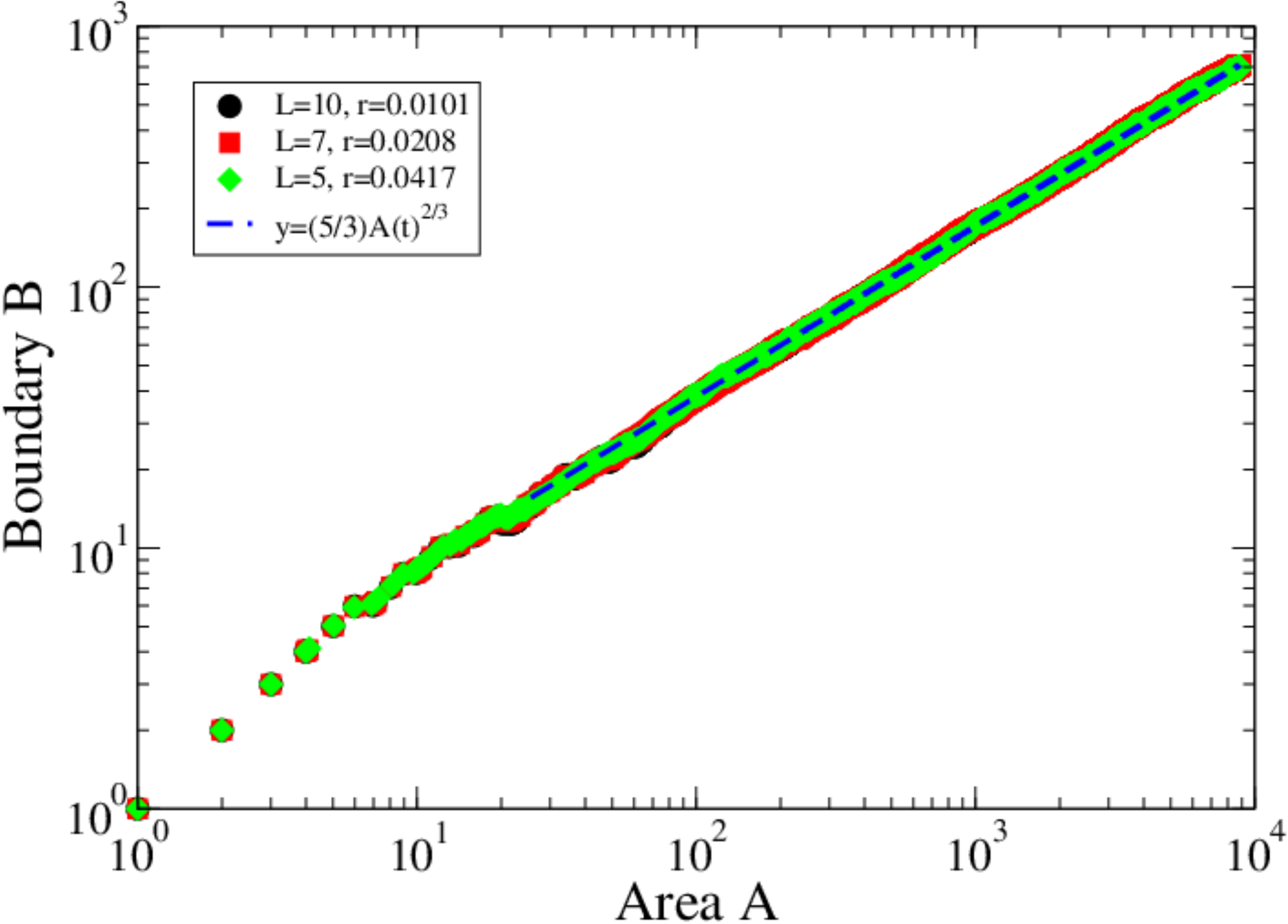}
        }
\caption{(color online) A plot (a) of the area $A(t)$ and the boundary $B(t)$ of sites visited by the particle for a specific CP initial configuration of rotators. For this particular configuration the particle's transient preparation stage is quite large, lasting until nearly time $t=10^5$. During this time $A(t)\propto t^{3/4}$ and $B(t)\propto t^{1/2}$ implying that particle's trajectory has fractal dimension $d_f\simeq 4/3$. In (b) a plot of $A(t)$ vs. $B(t)$ is shown for three different CP configurations where $r\simeq1/(\ell^2-1)$ for $\ell=5$, $7$, and $10$, respectively. For each of these configurations the particle's trajectory also has a fractal dimension of $d_f\simeq 4/3$, similar to that of a self-avoiding walk.}
\label{fig8}
\end{center}
\end{figure}

For different CP configurations the preparation stage varies from a few time steps to many thousands of steps. The particle's trajectory during this transient stage is fractal with a fractal dimension $d_f = 4/3$ for many of the different CP configurations we study (see in fig. \ref{fig8}(a)). Here, we calculated $d_f$ in two specific cases. The first is for a number of CP configurations that lead to propagation with an average time of $t\simeq 3\dot10^4$. The second is a number of non-propagating CP configurations for which propagation does not occur for $t\leq 10^5$. In both cases we find that the particle's trajectory has a fractal dimension $d_f\simeq 4/3$, as is shown in fig. \ref{fig8}(b). Moreover, the dynamics of the particle in both cases is subdiffusive with a TSD of $\bar \Delta(t) \propto t^{2/3}$.

Recall that the exponent $2/3$ and fractal dimension $d_f=4/3$ were also found in the CR and RP case for $r\simeq 0$. Therefore, when the initial configuration of rotators is nearly homogeneous the particle's transient behavior is nearly identical irrespective of whether the configuration is CR, RP, or CP. Moreover, during the particle's transient preparation stage the particle tries to randomize the initially periodic configuration as much as possible. (This is a possible explanation for the much slower subdiffusive dynamics observed during this time period.)

\begin{figure}[htbp]
\begin{center}
\includegraphics[scale=0.35]{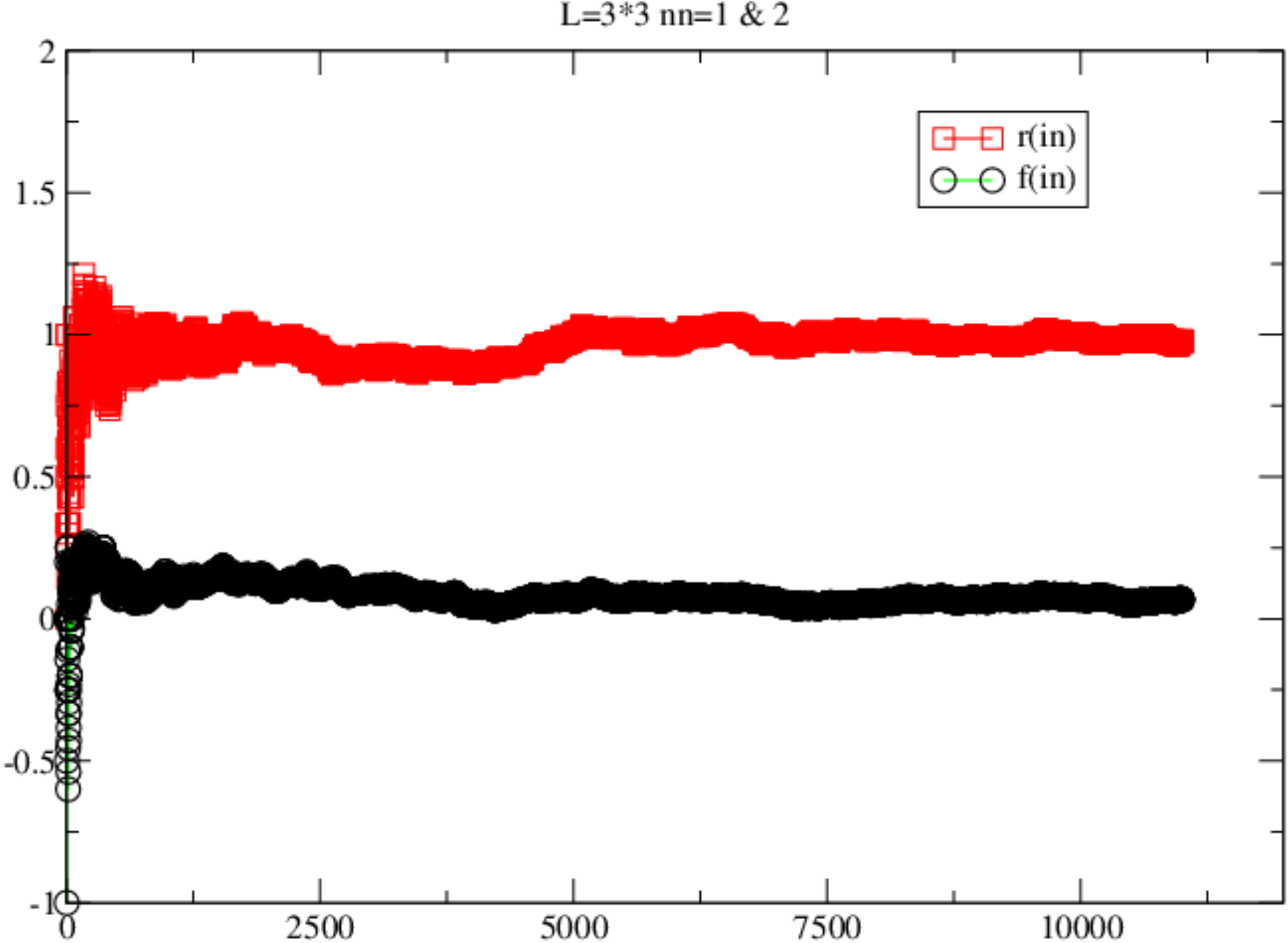}
\caption{A plot of the ratio $r_{in}(t)$ of left and right rotators (red squares) and the rotator-rotator correlation $F_{in}(t)$ (black circles) inside the sites visited by the particle for a specific CP configuration of rotators. Here, $r_{in}(t)$ approaches 1 and $F_{in}(t)$
approaches 0 as $t$ increases. Hence, as the particle moves through the lattice it makes the collection of rotators it visits both uncorrelated and evenly distributed.}
\label{fig9}
\end{center}
\end{figure}

To see this we calculate the rotator-rotator correlation $F_{in}(t)$ and ratio $r_{in}(t)$ inside the region visited by the particle by time $t$. We observe that $F_{in}(t)$ approaches 0 and $r_{in}(t)$ approaches 1 during the particle's transient preparation stage. Hence, as the particle moves through the lattice it tries to make the collection of rotators it encounters both uncorrelated and evenly distributed. An example of this is shown in fig. \ref{fig9} for a specific CP configuration.

\section{Discussion}\label{discussion}
We have studied both the transient and asymptotic behavior of a particle moving on a two-dimensional square lattice occupied by different initial configurations of right and left rotators. After scattering the particle these rotators change orientation so that the particle's motion influences the configuration of the rotator, which in turn influences the particle's motion. This interplay between the particle and the media or environment through which it moves is quite possibly one of the simplest such mechanisms one can study.

The particular configurations we have studied in this paper can be divided into three classes: completely random (CR), random periodic (RP), and completely periodic (CP). The first two of these configurations are random configurations in that they are generated via a specific random rule, whereas the third is deterministic. 

Besides distinguishing between these different types of configurations, we also study the extent to which the homogeneity of the configuration effects the particle's dynamics. This is done by varying the configuration's ratio $r$ of right to left rotators. For small $r\simeq 0$, where the configuration of rotators is nearly homogenous, one can consider the small minority of rotators with a left orientation to be defects or irregularities in the particle's environment, which the particle also creates as it moves through the lattice. Such defects can abruptly change the particle's motion. 

With this notion of defects in mind, we find that for the random initial configurations that we consider, i.e. CR and RP configurations, a small number of defects leads to subdiffusion whereas a large number of defects actually speeds up the particle's motion causing it to diffuse. Specifically, the particle's MSD for either a CR or RP configuration is numerically observed to be $\Delta(t)\propto t^{\alpha}$ where $\alpha$ increases monotonically from $\alpha\simeq2/3$ to $\alpha\simeq 1$ as $r$ increases from $r\simeq 0$ to $r\simeq 1$. Hence, we observe a transition in the particle's asymptotic dynamics from subdiffusion with an exponent of $\alpha=2/3$ to anomalous diffusion with an exponent of $\alpha=1$. Moreover, for $r\simeq 0$ the particle's trajectory has fractal dimension $d_f\simeq 4/3$, similar to a self-avoiding walk. For $r\simeq 1$, the trajectories fractal dimension $d_f\simeq 7/8$, which is the fractal dimension of a percolating cluster in 2-d \cite{pc}. The fact that these statistics do not depend on whether the configuration is CR or RP suggests that a periodically enforced ratio $r$ of right to left scatterers does not have a significant effect on the particle's dynamics. Rather we find that the difference in in whether the in initial configuration of rotators is randomly generated vs. deterministic.

One of the major differences between CR and RP configurations when compared with CP configurations is that, in the former the fastest motion through the lattice we observe is diffusion. Configurations that are CP, and therefore deterministic, typically cause the particle to propagate after some transient period of subdiffusion. This suggests that the randomness, i.e. disordered nature, of the CR and RP configurations ultimately impairs the particle's ability of developing an organized trajectory. In contrast, the far more structured CP configurations can cause the particle to have a much more organized trajectory, in which the particle eventually propagates in a single direction through the lattice.

Interestingly, before the particle begins to propagate for a given CP configuration, its motion is subdiffusive and in this sense very similar to the motion observed in the CR and RP cases for $r\simeq 0$. Based on this observation, it may be correct to think of the particle's transient preparation stage in a CP configuration as the time in which the particle is \emph{randomly searching} through the lattice for a way to propagate. The particle in the CR and RP case may also be searching for a way to propagate through the lattice. But because of the lack of structure in these configurations, the particle cannot find a way to do this and therefore can only (sub)diffuse through the lattice.

In contrast, one unifying feature of the particle's motion found over all configuration we study is that the particle in each case tries to make the orientation of the rotators it visits both uncorrelated and evenly distributed. This is shown by calculating the rotator-rotator correlation $F_{in}(t)$ and the time-dependent ratio $r_{in}(t)$ on the sites visited by the particle, which go in each case to 0 and 1, respectively.

Moving forward, it would be interesting to find a way to extend this study of a particle moving on a lattice to the more realistic case in which the particle moves in continuous space. Specifically, it would be interesting to study the how the particle's dynamics in the continuous setting would be effected by both random as well as periodic defects. Another question to consider is how the particle's dynamics on the square and other lattices would be effected by such defects if some of the lattice sites were vacant, so that the particle could pass through a lattice site without its velocity being effected.

\begin{acknowledgments}
Shradha Mishra and Sanchari Bhattacharya would like to acknowledge  DST INSPIRE faculty award 2012 for financial support.
\end{acknowledgments}



\begin{thebibliography}{20}
\bibitem{Lorentz} The Lorentz in Lorentz lattice gas refers to H. A. Lorentz who assumed that the electrons in a conductor move independently of each other so that it is sufficient to study a single electron.
\end{thebibliography}
\end{document}